\documentclass[twocolumn]{aastex7}

\newcommand{\Msun}{M$_{\odot}$}
\newcommand{\Mbh}{$M_{\rm BH}$}

\newcommand{\Mstar}{$M_{\star}$}

\newcommand{\Lsun}{L$_\odot$}

\newcommand{\ml}{\emph{M/L}}

\newcommand{\hst}{\emph{HST}}

\newcommand{\kms}{km~s$^{-1}$}

\newcommand{\go}{G$_{\rm outer}$}

\newcommand{\cotwo}{$^{12} $CO(2$-$1)}

\newcommand{\siggas}{$\sigma_{\rm gas}$}

\newcommand{\kinms}{\textsc{KinMS}}
\newcommand{\skysampler}{\textsc{SkySampler}}

\hypersetup{colorlinks, linkcolor=blue, citecolor=cyan, urlcolor=magenta}

\usepackage{multirow}
\usepackage{footnote}
\usepackage{newtxtext,newtxmath}
\usepackage{graphicx}	
\usepackage{amsmath}	
\usepackage{lineno}
\usepackage{caption}
\usepackage{xcolor}
\linenumbers

\graphicspath{{./}{figures/}}
\received{January 26, 2026}
\submitjournal{AAS Journals}

\shorttitle{Evidence for a Massive Black Hole in NGC~4061}
\shortauthors{D.\ D.\ Nguyen, L.\ Q.\ T.\ Nguyen $\&$ H.\ N.\ Ngo et al.}
\begin{document}
\title{Dynamical Evidence for a Billion Solar Mass Black Hole in Galaxy NGC~4061 from ALMA \cotwo\ Kinematics}

\correspondingauthor{Dieu D.\ Nguyen} \email{dieun@umich.edu}
\author[0000-0002-5678-1008]{Dieu D.\ Nguyen}
\email{dieun@umich.edu}
\affiliation{Department of Astronomy, University of Michigan, 1085 South University Avenue, Ann Arbor, MI 48109, USA}
\author[0009-0006-4602-1968]{Long Q.\ T.\ Nguyen}
\email{baobi123987@gmail.com}
\affiliation{Faculty of Physics—Engineering Physics, University of Science, Vietnam National University in Ho Chi Minh City, Vietnam}
\author[0000-0001-5802-6041]{Elena Gallo}
\email{egallo@umich.edu}
\affiliation{Department of Astronomy, University of Michigan, 1085 South University Avenue, Ann Arbor, MI 48109, USA}
\author[0009-0006-5852-4538]{Hai N.\ Ngo}
\affiliation{Faculty of Physics—Engineering Physics, University of Science, Vietnam National University in Ho Chi Minh City, Vietnam}
\email{hai10hoalk@gmail.com}
\author[0000-0001-9649-2449]{Que T.\ Le}
\email{ltque@hcmiu.edu.vn}
\affiliation{Department of Physics, International University, Vietnam National University in Ho Chi Minh City, Vietnam}
\author[0000-0001-9879-7780]{Fabio Pacucci}
\email{fabio.pacucci@cfa.harvard.edu}
\affiliation{Center for Astrophysics—Harvard $\&$ Smithsonian, 60 Garden St., Cambridge, MA 02138, USA}
\affiliation{Black Hole Initiative, Harvard University, 20 Garden St., Cambridge, MA 02138, USA}
\author[0009-0004-3689-8577]{Tinh Q.\ T.\ Le}
\email{lethongquoctinh01@gmail.com}
\affiliation{Department of Physics, International University, Vietnam National University in Ho Chi Minh City, Vietnam}
\author[0009-0009-0015-1208]{Tuan N.\ Le}
\affiliation{Faculty of Physics—Engineering Physics, University of Science, Vietnam National University in Ho Chi Minh City, Vietnam}
\email{tuan.le.nutshell@gmail.com}
\author[0009-0005-8845-9725]{Tien H.\ T.\ Ho}
\email{htien2808@gmail.com}
\affiliation{Faculty of Physics—Engineering Physics, University of Science, Vietnam National University in Ho Chi Minh City, Vietnam}

\begin{abstract}

We present the first robust dynamical measurement of the supermassive black hole (SMBH) mass in the massive early-type galaxy NGC~4061 using high–spatial-resolution ALMA observations of the \cotwo\ emission. By combining archival Cycle~6 data with new Cycle~7 observations, we achieve a synthesized beam of $0\farcs16 \times 0\farcs13$, comparable to the expected sphere of influence of the central black hole. The molecular gas forms a regularly rotating circumnuclear disk aligned with the prominent dust lane seen in HST imaging. We model the full three-dimensional ALMA data cube using the \kinms\ forward-modeling framework, exploring both data-driven and analytic prescriptions for the gas surface brightness distribution. Our Bayesian analysis yields a best-fitting SMBH mass of $M_{\rm BH} = (1.17^{+0.08}_{-0.10}\,[{\rm stat.}] \pm 0.43\,[{\rm syst.}]) \times 10^{9}$~\Msun\ and an $I$-band stellar mass-to-light ratio of $M/L_{\rm F814W} = 3.46^{+0.07}_{-0.06}\,[{\rm stat.}] \pm 0.10\,[{\rm syst.}]$~\Msun/\Lsun. The inferred black hole mass is fully consistent across different modeling assumptions and remains insensitive to plausible radial variations in the $M/L_{\rm F814W}$ profile. Our results resolve the long-standing discrepancy between previous indirect mass estimates based on conflicting stellar velocity dispersion measurements and demonstrate that the exceptionally large dispersion reported in the literature is likely spurious. This study highlights the power of high-resolution ALMA molecular gas kinematics for precision SMBH mass measurements at the high-mass end of the local black hole mass function.

\end{abstract}



\section{Introduction}\label{sec:intro}

\begin{figure*}
    \centering
    \includegraphics[width=2\columnwidth]{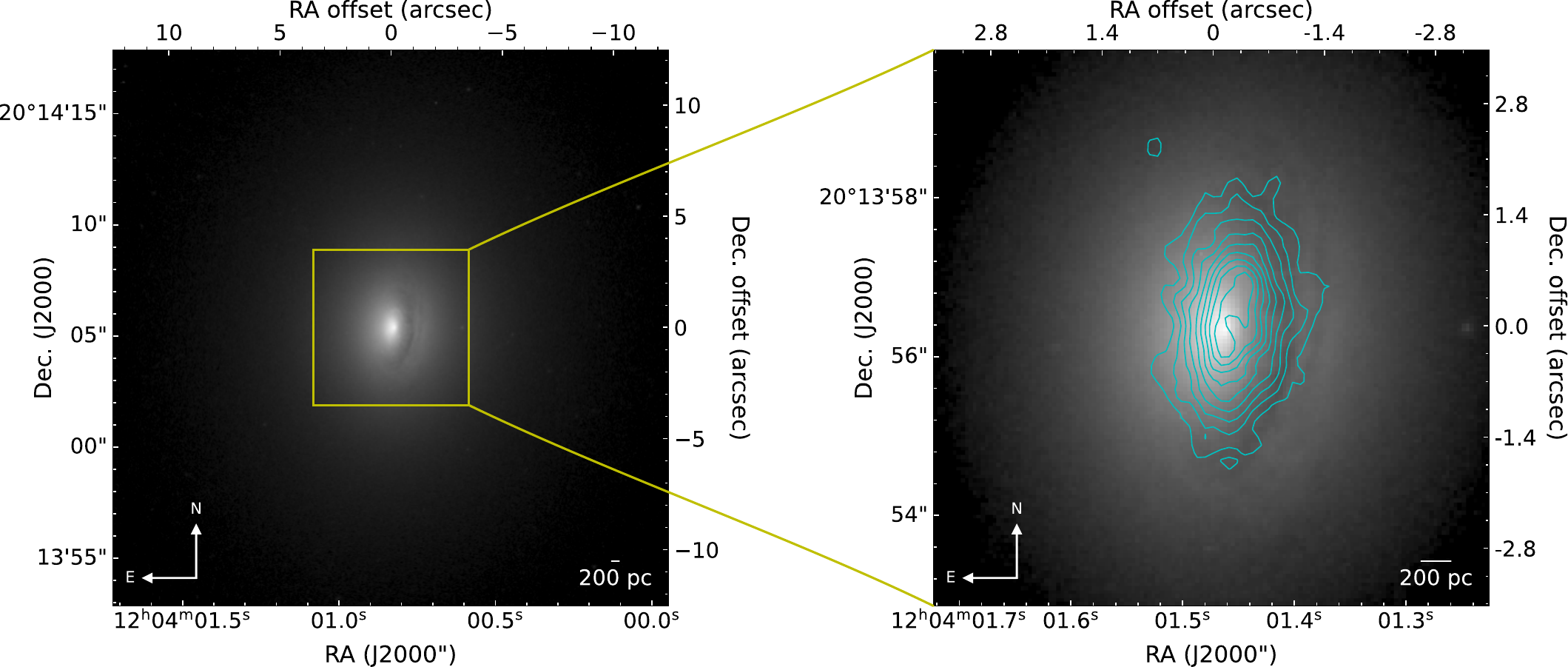}
    \caption{{\it Left:} HST/WFPC2 F814W image of NGC~4061 within a 25\arcsec~$\times$~25\arcsec\ (or 13~kpc~$\times$~13~kpc) field of view, showing the central dust lane. {\it Right:} ALMA \cotwo\ integrated intensity contours overlaid on the left zoom-in HST image (7\arcsec~$\times$~7\arcsec\ or 3.64~kpc~$\times$~3.64~kpc), illustrating the alignment of the molecular gas with the dust lane.}
    \label{fig:HST-overlaid}
\end{figure*}

Supermassive black holes (SMBHs) are widely recognized as key drivers of galaxy formation and evolution \citep{ferrarese2000fundamental,  mcconnell2013revisiting, Saglia16}. This understanding is supported by well-established correlations between SMBH mass (\Mbh) and large-scale host galaxy properties such as bulge luminosity \citep{Magorrian98} and stellar velocity dispersion \citep[$\sigma$;][]{ferrarese2000fundamental}. These relations even have broader implications, for examples:  (i) constraining models of black hole (BH) fueling and feedback \citep{Silk98, Fabian12, Netzer15, Naab17}, (ii) defining the SMBH mass function relevant for gravitational-wave background estimates for Pulsar Timing Arrays \citep{Kelley25} and space-based \citep[LISA;][]{Postiglione25} detectors, and (iii) providing the calibration benchmark for reverberation-mapped active galactic nuclei (AGN) used to determine the virial factor in single-epoch \Mbh\ estimations \citep{Reines13, Greene24}. Moreover, these local relations serve as a baseline for testing potential redshift evolution in the SMBH–galaxy connection \citep{Nguyen23, Nguyen2025b, Nguyen2025d, Pacucci23} and for constraining the nature of BH seeds at $z > 15$ \citep[e.g.,][]{Greene2020, Inayoshi2020}. 

However, these SMBH–host galaxy relations remain poorly constrained at both the low-mass \citep[$M_{\rm BH}\lesssim10^6$~\Msun;][]{Nguyen14, Nguyen17conf, Nguyen2025b} and high-mass \citep[$M_{\rm BH}\gtrsim10^9$~\Msun;][]{Thomas2016, Mehrgan2019, Nightingale2023, Melo-Carneiro2025} ends. These relations are more complex than initially assumed, as galaxies with diverse structural properties \citep{Cappellari16, Graham2025} and evolutionary histories—such as brightest cluster galaxies \citep{McConnell2012, Krajnovic18a}, massive core ellipticals \citep{McConnell2011, Graham2013}, compact early-type systems \citep{Seth14, NGuyen17, Nguyen18, Nguyen19, Ahn18, Voggel18, Taylor2015}, and low-mass spirals \citep{denBrok15, Baldassare15, Nguyen22}—exhibit notable deviations from the canonical scaling relations \citep{Hlavacek-Larrondo2012, vandenbosh2016}. To refine our understanding of the coevolution between SMBHs and their hosts, additional robust \Mbh\ measurements are essential across the full mass spectrum and for a broad range of galaxy types \citep{Greene20}.

Located in the NGC~4065 group of galaxies, NGC~4061 (also catalogued as NGC~4055; R.A. = $12^{\rm h}04^{\rm m}01\fs4569$, Decl. = ${+}20\degr 13\arcmin 56\farcs 470$) has been studied primarily as a bent-tail radio source, whose morphology reflects its motion through the intragroup medium and interaction with the surrounding environment \citep{Doe1995}. These environmental signatures highlight both the richness of its dynamical context and the possibility that the central region may host significant kinematic complexity \citep{Freeland2010}. The galaxy thus presents an intriguing but underexplored laboratory for expanding the SMBH census toward the upper end of the mass distribution and for testing \Mbh\ measurement techniques.

The Hypercat catalog lists an exceptionally large stellar velocity dispersion of $\sigma\approx459$~\kms, implying an extremely massive BH ($M_{\rm BH} \approx 1.2\times10^{10}$~\Msun) from the \citet{Kormendy13} relation. However, \citet{Pinkney2005} reported a much lower central dispersion of $\sigma\approx290$~\kms, based on stellar kinematics derived long slit on the LDSS-3 instrument at the Magellan Clay telescope, suggesting a more moderate BH mass of \Mbh~$\approx1.4\times10^9$~\Msun\ and indicating that the earlier value was likely overestimated. Additionally, optical observations from HST reveal a well-defined dust disk with a radius of 2\farcs5 (left panel of Figure~\ref{fig:HST-overlaid}), while ground-based spectra show central H$\alpha$ emission from rapidly rotating gas, with a velocity gradient of 270~\kms\ across 0\farcs55 \citep{Pinkney2005}. This disk is nearly edge-on and closely aligned with the galaxy’s projected major axis.

\begin{table*}
	\centering
	\caption{Properties of ALMA observing tracks.}
    \vspace{-3mm}
	\label{tab:Observing Tracks}
	\begin{tabular}{cccccccccc}
	\hline\hline
Project code&Obs. tracks&Obs. date&Config.&Baseline range&ToS&   MRS   &$\theta_{\rm FWHM}$&Calibration\\
            &     &         &       &              &(seconds)&(\arcsec)&(\arcsec)&Pipeline\\
     (1)    & (2) &    (3)  &  (4)  &      (5)     &   (6) &     (7)     & (8) &  (9)    \\ 
	\hline
\multirow{2}{*}{2018.1.00397.S}&{\tt uid\_A002\_Xe03886\_X7606}&2019-08-19&C43-7&41.4~m--3.2~km&2780&\multirow{2}{*}{1.6}&\multirow{2}{*}{0.112}&\multirow{2}{*}{\textsc{CASA}~5.4.0-70}\\
                                                  &{\tt uid\_A002\_Xe03886\_Xd75e}&2019-08-20&C43-7&41.4~m--3.4~km&2805&  &  &\\
    \hline
     2019.1.00036.S                   &{\tt uid\_A002\_Xea90c0\_X334e}  &2021-03-29&C43-5& 15.0~m--1.3~km &3731& 4.3&0.317&\textsc{CASA}~6.1.1.15\\ \hline
	\end{tabular}
    \parbox[t]{\textwidth}{\small \textbf{Notes:} Columns: (2) Observational track ID. (3) Observational date (year-month-day). (4) ALMA configuration array. (5) Minimum and maximum baseline length. (6) Total on-source time (ToS). (7) Maximum recoverable scale (MRS), i.e.\ the largest angular scale that can be recovered with the given array. (8) the average synthesized beam size. (9) Calibration method.} 
\end{table*}

\begin{table*}
	\centering
	\caption{Properties of spectral windows of adopted observing tracks.}
    \vspace{-3mm}
	\label{tab:Spectral windows}
	\begin{tabular}{ccccccccc}
		\hline\hline
Project code & Obs. track & SPW & Target &Freq. range&Cen. freq.&Flagged&Bandpass& Phase\\
             &       &    &(Line)&     (GHz) &  (GHz)  &    & (Quasars) & (Quasars)\\
         (1) &  (2)  & (3) &    (4)    &          (5)         &     (6)     &       (7)       &(8) & (9)\\ 
		\hline
\multirow{4}{*}{2018.1.00397.S}&\multirow{4}{*}{{\tt uid\_A002\_Xe03886\_X7606}}&0&\cotwo&(224.09, 225.96)&225.0 &No&\multirow{4}{*}{J1215+1654}&\multirow{4}{*}{J1157+1638}\\ 
  	          &                                                &1& cont. &(225.98, 227.97)&227.0&No& &\\ 
                  &                                                &2& cont. &(239.65, 241.63)&240.6&No& &\\ 
                  &                                                &3& cont. &(241.56, 243.58)&242.6&No& &\\
        \hline
\multirow{4}{*}{2018.1.00397.S}&\multirow{4}{*}{{\tt uid\_A002\_Xe03886\_Xd75e}}&0&\cotwo&(224.09, 225.96)&225.0&No&\multirow{4}{*}{J1215+1654}&\multirow{4}{*}{J1157+1638}\\ 
  	          &                                                &1& cont. &(225.98, 227.97)&227.0&No& &\\ 
                  &                                                &2& cont. &(239.65, 241.63)&240.6&No& &\\ 
                  &                                                &3& cont. &(241.56, 243.58)&242.6&No& &\\
        \hline        
\multirow{4}{*}{2019.1.00036.S}&\multirow{4}{*}{{\tt uid\_A002\_Xea90c0\_X334e}} &0& cont. &(222.20, 224.07)&223.1&No&\multirow{4}{*}{J1058+0133}&\multirow{4}{*}{J1215+1654}\\ 
  	          &                                                &1&\cotwo&(224.19, 226.06)&225.1&No& &\\ 
                  &                                                &2& cont. &(236.12, 237.99)&237.1&No& &\\ 
                  &                                                &3& cont. &(238.31, 240.19)&239.3&No& &\\\hline
	\end{tabular} \\
\parbox[t]{\textwidth}{\small \textbf{Notes:} Columns: (1) Project code. (2) Observational track ID. (3) SPW ID. (4) Targeted line of each SPW. (5) Frequency range of each SPW. (6) Central frequency of each SPW. (7) Whether the SPW is flagged manually. (8)--(9) Quasars used to do bandpass and phase calibration.} 
\end{table*}

\begin{table}
\centering
	\caption{Parameters of the 1.3~mm combined continuum image and source.}
	\label{tab:continuum}
    \vspace{-3mm}
	\begin{tabular}{lc} 
	\hline\hline
Image property                & Value           \\
	\hline
Image size (pixel$^2$)        & $512\times512$  \\
Image size (arcsec$^2$)       & $20.5\times20.5$\\
Image size (kpc$^2$)          & $10.7\times10.7$\\
Pixel scale (arcsec per pixel)& 0.04            \\
Pixel scale (pc per pixel)    & 20.8            \\
Sensitivity (mJy per beam)    & 0.02            \\
Synthesised beam (arcsec$^2$) & $0.16\times0.13$\\
Synthesised beam (pc$^2$)     & $83.2\times67.6$\\
	\hline
Source property               & Value           \\
	\hline
Right ascension               & $12^{\rm h}04^{\rm m}01\fs4569$   \\
Declination                   & ${+}20\degr13\arcmin56\farcs470$  \\
Integrated flux (mJy)         & $3.48\pm0.18$                     \\
Deconvolved size (arcsec$^2$) & $(0.26\pm0.03)\times(0.22\pm0.03)$\\
Deconvolved size (pc$^2$)     & $(135.2\pm 15.6)\times(114.4\pm15.6)$\\
	\hline
	\end{tabular}
\end{table}

\begin{table}
\centering
    \caption{The combined \cotwo\ data cube properties}
    \label{tab:co}
    \begin{tabular}{lr}
    \hline\hline
 CO image property                    & Value              \\ \hline
Spatial extent (pixel$^2$)           & 512 $\times$ 512   \\ 
Spatial extent (arcsec$^2$)        & 20.5 $\times$ 20.5 \\ 
Spatial extent (kpc$^2$)             & 10.7 $\times$ 10.7 \\ 
Pixel scale (arcsec per pixel)      & 0.04               \\ 
Pixel scale (pc per pixel)             & 20.8               \\ 
Velocity range (km s$^{-1}$)       & 6815 -- 7550       \\ 
Channel width (km s$^{-1}$)       & 10                 \\ 
Number of constraints                 & 78,100             \\
Mean synthesised beam (arcsec$^2$)  & 0.16 $\times$ 0.13 \\ 
Mean synthesised beam (pc$^2$)        & $83.2\times67.6$   \\ 
Sensitivity (mJy per beam per 10 \kms)& 0.4                \\
    \hline
    \end{tabular}
\parbox[t]{0.47\textwidth}{\small \textbf{Notes:} Assumed center of the \cotwo\ circumnuclear disk, coincident with the 1.3~mm continuum emission described in Table~\ref{tab:continuum}.} 
\end{table}

\begin{figure*}
    \centering
    \includegraphics[width=0.88\textwidth]{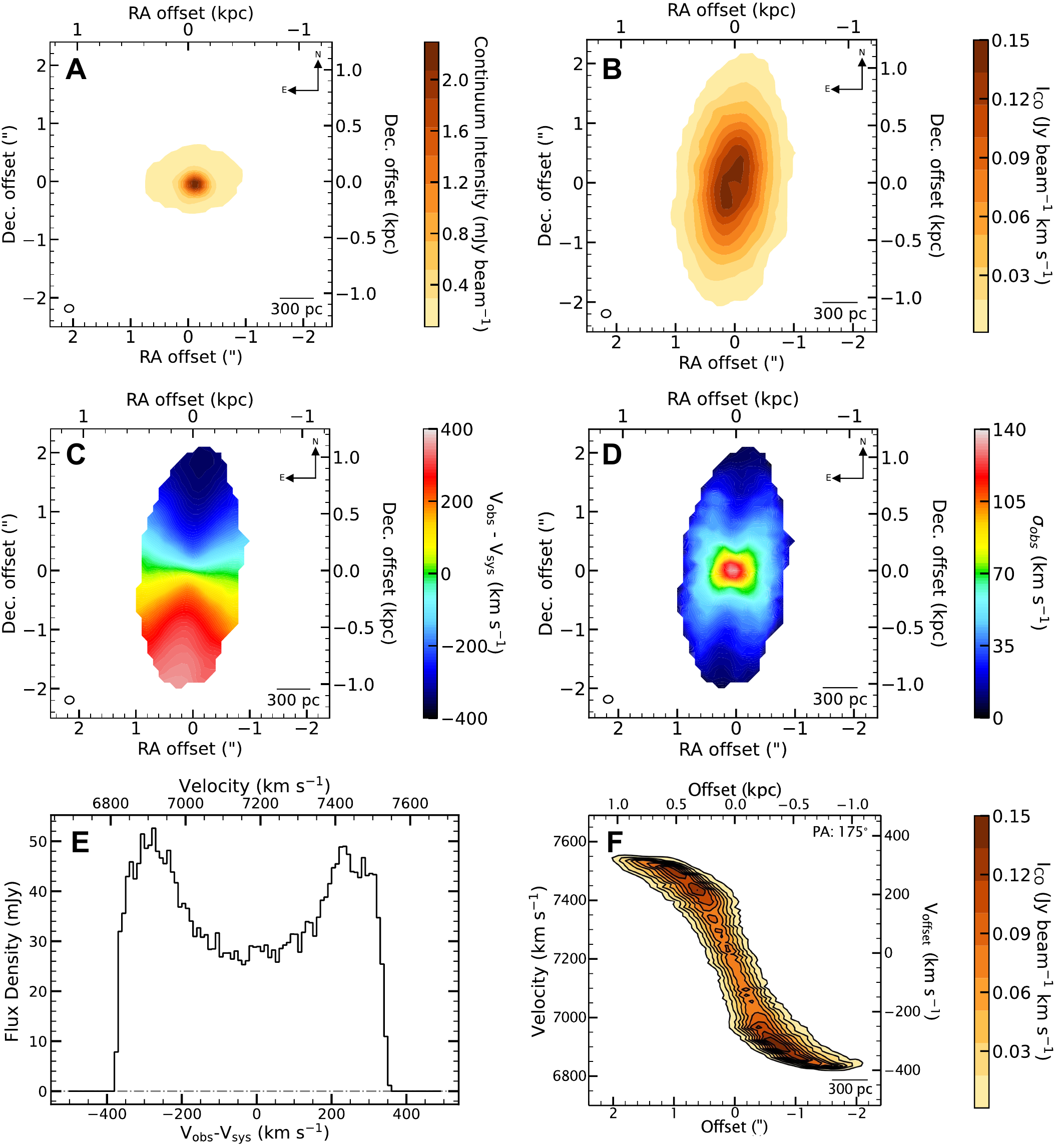}
    \caption{\textit{Panel A}: 1.3~mm continuum emission in Band~6. \textit{Panels B–D}: ALMA \cotwo\ moment maps of NGC~$4061$—integrated intensity, intensity-weighted mean LOS velocity, and intensity-weighted LOS velocity dispersion, respectively. The synthesized beam is shown as a black ellipse in the lower-left corner of each map. \textit{Panel E}: Integrated spectrum extracted from a $6\arcsec \times 6\arcsec$ ($3.12 \times 3.12$~kpc) box; the horizontal dot-dashed line marks zero flux. \textit{Panel F}: Position–velocity diagram along the major axis, adopting a systemic velocity of $v_{\rm sys}=7190$~\kms\ and position angle $\Gamma=175\degr$.}
    \label{fig:moment-maps}
\end{figure*}

NGC~4061 is a massive early-type galaxy (ETG) at a distance of $D = 107.2$~Mpc, with an effective radius estimated in the range $R_e \sim 4.2$--$9.1$~kpc and a $K$-band absolute magnitude of $M_K=-25.3$~mag \citep{Ma14}. Using the prescription from equation~(2) of \citet{Cappellari2013a}, calibrated from 260 ATLAS$^{\rm 3D}$ ETGs: $\log(M_\star)= 10.58-0.44\times(M_K+23)$, we estimate a stellar mass of \Mstar\ $\approx 4.3 \times 10^{11}$~\Msun, placing NGC~4061 near the high-mass end of the \Mbh--$\sigma$ relation and reinforcing the likelihood that it hosts a SMBH.

Despite these strong dynamical signatures, no published, peer-reviewed, high-precision \Mbh\ measurement yet exists for NGC~4061. The geometry of its dust disk and the observed gas rotation provide a compelling case for high-spatial-resolution dynamical modeling (stellar or gas) to refine the mass estimate and test for systematic biases. Moreover, the galaxy’s group environment, bent-tail radio morphology, and potential AGN activity underscore the importance of linking central BH growth with its environmental and structural context.

In this work, we present new high-sensitivity, high-spatial-resolution observations of the \cotwo\ circumnuclear disk (CND) in the central region of NGC~4061 obtained with the Atacama Large Millimeter/submillimeter Array (ALMA). These data allow us to map the cold molecular gas kinematics and, for the first time, to dynamically constrain the mass of its central SMBH precisely.

In Section~\ref{alma}, we describe the ALMA observations, data reduction, imaging, light-of-sight (LOS) kinematic measurements, and molecular gas mass derivation from the \cotwo\ molecular gas. Section~\ref{sec:stellar-mass} presents the stellar mass modeling of NGC~4061 based on HST data, respectively. The molecular gas dynamical modeling used to measure \Mbh\ and assess associated uncertainties are detailed in Section~\ref{sec:dynamical-models}, followed by conclusions in Section~\ref{sec:conclusion}. Given our adopted flat $\Lambda$CDM cosmological model with $H_0 = 70$~\kms~Mpc$^{-1}$, $\Omega_{\Lambda,0} = 0.7$, and $\Omega_{\rm m,0} = 0.3$, and a distance of $D = 107.2$~Mpc \citep{Ma14}, the physical scale is 520~pc~\arcsec$^{-1}$.

\section{ALMA Observations}\label{alma}

\subsection{Data Reduction}\label{reduction}

We used the ALMA \cotwo\ circumnuclear disk (CND) observations of NGC~4061 obtained during Cycles~6 (PID: 2018.1.00397.S; PI: M. Smith) and 7 (PID: 2019.1.00036.S; PI: D. Nguyen). The data were taken with the 12-m array in the C43-7 and C43-5 configurations, respectively, providing a primary-beam full width at half maximum (FWHM) of $\approx$25\arcsec. A summary of these observing tracks is given in Table~\ref{tab:Observing Tracks}.

For each execution block, the observations were carried out in Band~6 using four frequency-division-mode (FDM) spectral windows (SPWs). Each SPW spans 2~GHz and is divided into 1920 channels (976.562~kHz, corresponding to $\approx$1.3~\kms). One SPW was centered on the \cotwo\ transition ($v_{\rm rest} = 230.538$~GHz), while the remaining three SPWs were allocated to measure the continuum. The detailed spectral setup is listed in Table~\ref{tab:Spectral windows}.

Calibration was carried out using the Common Astronomy Software Applications package, \textsc{CASA}\footnote{\url{https://casa.nrao.edu/}} \citep{McMullin07}, with version~5.4.0-70 for the Cycle~6 data and version~6.1.1.15 for the Cycle~7 data. All datasets were processed using the standard ALMA Science Pipeline\footnote{\url{https://almascience.eso.org/processing/science-pipeline}} to obtain three measurement sets (MS) for three observational tracks listed in Tables~\ref{tab:Observing Tracks} and \ref{tab:Spectral windows}.

We combined the visibility data from all three MS into a combined and calibrated MS using the \texttt{concat} task in \textsc{CASA}. The optimal visibility–weight scaling factors applied in the \texttt{visweightscale} mode were 0.6 for the \texttt{uid\_A002\_Xea90c0\_X334e} MS (2019.1.00036.S), and 0.2 for each of the \texttt{uid\_A002\_Xe03886\_X7606} and \texttt{uid\_A002\_Xe03886\_Xd75e} MSs (2018.1.00397.S).

\subsection{Creating the Combined 1.3 mm Continuum Image}\label{sec:cont}

We produced the combined continuum image at 1.3mm using the \textsc{CASA} {\tt tclean} task in multifrequency synthesis mode \citep{Rau11}, combining the continuum SPWs with line-free channels from the targeted SPWs of all three MS. Briggs weighting with a robust parameter of 0.5 was applied to optimize the balance between sensitivity and resolution. The resulting image shows a resolved source with a size of $1\farcs8\times1\farcs2$ that has an RMS noise of $\sigma_{\rm cont} = 120~\mu$Jy~beam$^{-1}$ and a synthesized beam of $\theta_{\rm FWHM,cont} = 0\farcs 16 \times 0\farcs 13$ at a position angle of $\Gamma=352.9$\degr. 

Figure~\ref{fig:moment-maps} (Panel~A) displays the continuum image, showing a single source that matches the galaxy’s kinematic center (best-fitting SMBH position; see Section~\ref{sec:bayesian-infer}, and also defined as the galaxy center), with an integrated flux density of $3.48 \pm 0.18$~mJy, consistent within the $\approx$10\% ALMA flux calibration uncertainty. A two-dimensional (2D) Gaussian fit using the {\tt imfit} routine in \textsc{CASA} confirms the source is spatially resolved. Continuum image parameters and source properties are summarized in Table~\ref{tab:continuum}.

\subsection{Creating the Combined-\cotwo\ Data Cube}\label{sec:line}

Given that the 1.3 mm continuum emission is detected at the nucleus of NGC 4061 and is spatially resolved in all three MSs summarized in Table \ref{tab:Spectral windows}, we used the combined and calibrated MS (Section~\ref{reduction}) with all line-free channels across the 12 targeted SPWs (Table~\ref{tab:Spectral windows}). The continuum was modeled with a linear power-law function and subtracted from the combined and calibrated visibilities in the \textit{uv} plane to isolate the \cotwo\ line emission. This subtraction was carried out using the \textsc{CASA} task {\tt uvcontsub} (e.g., \citealt{Davis20, Nguyen20}).

We produced the final three-dimensional (R.A., Decl., velocity) cube using the \texttt{tclean} task in \textsc{CASA}. To model the \cotwo\ CND, constrain the mass of the central compact object, and optimize the balance between surface-brightness sensitivity and spatial resolution, we adopted a Högbom deconvolver \citep{Hogbom74}. The resulting cube has dimensions of 512~$\times$~512 pixels$^2$ with a pixel scale of $0\farcs04$, which adequately samples the synthesized beam while maintaining a manageable file size. We used a channel width of 10~\kms—an optimal choice for SMBH dynamical modeling \citep{Davis14, Nguyen21, Ngo2025a}—and Briggs weighting with a robust parameter of 0.5. The velocity axis was referenced to the rest frequency of 230.538~GHz.

During interactive imaging, we applied a clean mask to suppress sidelobe artifacts and improve the fidelity of the recovered emission. The continuum-subtracted dirty cube was cleaned down to a threshold of 1.5 times the root-mean-squared (RMS) noise level ($\sigma_{\rm RMS}$; measured from line-free channels). The final self-calibrated, cleaned cube shares the same synthesized beam as the 1.3~mm continuum image (Section~\ref{sec:cont}), and its properties are summarized in Table~\ref{tab:co}.

\begin{table}
\centering
\caption{Gas MGE model}
\begin{tabular}{cccc} 
\hline\hline
$j$&$\lg\Sigma_{\rm ISM,}$$_j$ (\Msun\ ${\rm pc^{-2}})$&$\lg\sigma_j(\arcsec)$&$q_j=b_j/a_j$\\
(1) & (2) & (3) & (4)\\ 
\hline
1 & 1.98 & $-$0.23 & 0.94 \\
2 & 2.71 & $-$0.09 & 0.50 \\
3 & 1.17 &    0.32 & 1.00 \\
\hline
\end{tabular}
\parbox[t]{0.47\textwidth}{\small \textit{Notes:} (1) the Gaussian component, (2) the luminosity surface density, (3) the Gaussian dispersion along the major axis, and (4) the axial ratio.} 
\label{table:mge-ism}
\end{table}

\subsection{\cotwo\ Emission Moment Maps}\label{sec:momentmaps}

The \cotwo\ emission spans in a velocity range of 6815--7550~\kms, with a systemic velocity of $v_{\rm sys}\approx 7190$~\kms. Moment maps, including the zeroth (integrated intensity; panel~B), the first (intensity-weighted mean light-of-sight (LOS) velocity; panel~C), and the second (intensity-weighted LOS velocity dispersion; panel~D), were derived from the \cotwo\ data cube using the masked-moment method \citep{Dame11}, as shown in Figure~\ref{fig:moment-maps}.

We first produced a smoothed version of the original data cube by duplicating it and applying a Gaussian spatial convolution with $\sigma = 1.5 \times \theta_{\rm FWHM}$ and spectral smoothing with a Hanning window four times the channel width \citep{Smith21, Dominiak24} to that copy of the original data cube. A mask was then generated by applying a $0.5\sigma_{\rm RMS}$ threshold to the unsmoothed cube (equivalent to $8\sigma_{\rm RMS}$ in the smoothed cube). This method effectively suppresses noise while retaining most of the flux. Pixels that exceeding the threshold in the smoothed cube (or the mask) were used to create moment maps from the unsmoothed cube.  

The zeroth-moment map shows that the CND extends $\approx$4\arcsec\ along the major axis and 2\arcsec\ along the minor axis, featuring a smooth intensity gradient and peaks at the CND's center. Its also illustrates the alignment between the \cotwo\ emission and the dust disk (see Figure~\ref{fig:HST-overlaid}).  The total molecular gas mass was estimated by using the ``CO-to-H$_2$ conversion factor'' of $X_{\rm CO} = 2 \times 10^{20}{\rm cm^{-2}(K\,km~s^{-1})^{-1}}$ \citep[or $\alpha_{\rm CO} = 4.3$ \Msun\ (K\,\kms\ pc$^{-1}$)$^{-1}$;][]{Bolatto2013}:
\begin{equation*} 
\small
	\begin{split} 
M_{\rm gas} &= 1.05 \times 10^4\left( \frac{X_{\rm CO}}{2 \times 10^{20} \dfrac{\rm cm^{-2}}{\rm K\ km\ s^{-1}}} \right) \left( \frac{1}{1+z} \right) \left( \frac{S_{\rm CO}\Delta v}{\rm Jy\ km\ s^{-1}} \right) \left( \frac{D_L}{\rm Mpc} \right)^2, 
	\end{split} 
\end{equation*}
where $S_{\rm CO} \Delta v = 4.2$ Jy \kms\ is the integrated flux density derived from our data cube. For NGC~4061 ($z \approx 0.024450$; NASA/IPAC Extragalactic Database\footnote{NED: \url{https://ned.ipac.caltech.edu}}), we adopted a luminosity distance of $D_L = 107.2$~Mpc \citep{Ma14} and a flux density ratio of unity between $^{12}$CO(2–1) and $^{12}$CO(1–0) \citep{Smith21}. Under these assumptions, the total molecular gas mass is $M_{\rm gas} \approx 4.95 \times 10^8$~\Msun.

\begin{figure}
    \centering
    \includegraphics[width=0.49\textwidth]{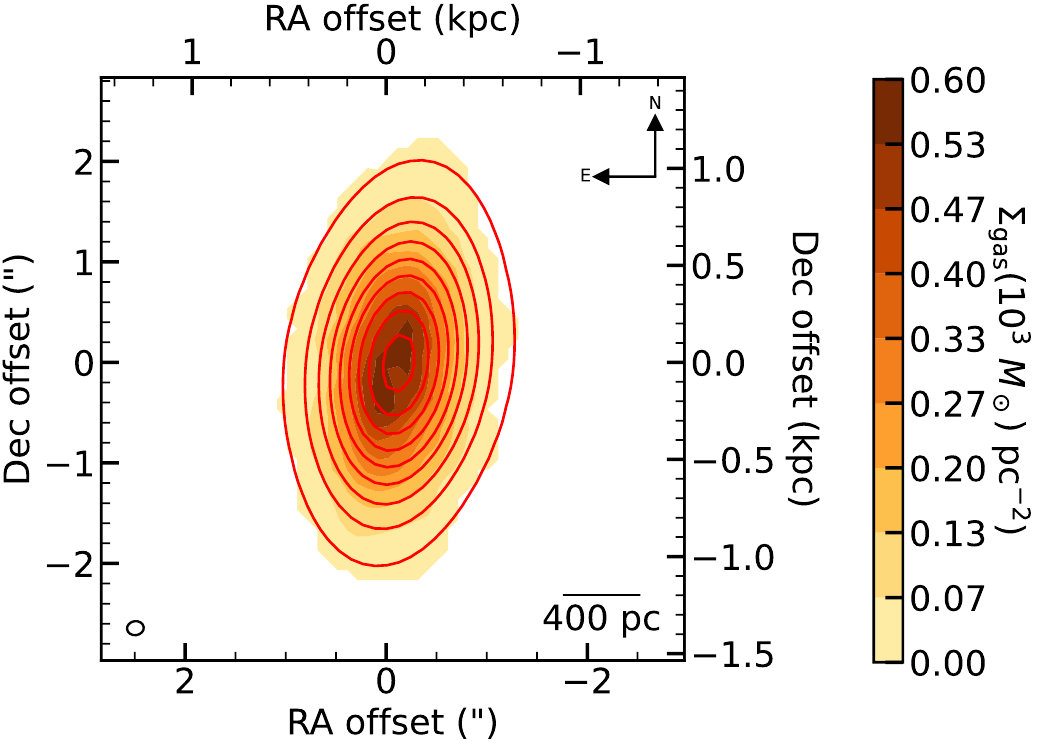}
    \caption{Comparison between the molecular gas mass distribution derived from the zeroth-moment map (panel~B of Figure~\ref{fig:moment-maps}) and its MGE model (contours). The strong correspondence between the data and model demonstrates excellent agreement across matching radii and contour levels.}
    \label{fig:CO_mge}
\end{figure}

The first-moment map shows a regularly intensity-weighted mean LOS rotating, unwarped thin disk with velocities up to $\pm370$~\kms. The second-moment map indicates moderate turbulence, with a gradient in the intensity-weighted LOS velocity dispersion of $0 \lesssim \sigma_{\rm LOS} \lesssim 70$~\kms\ outside the central boxy region ($0\farcs5 \times 0\farcs5$). Within this region, $\sigma_{\rm LOS}$ rises to $\approx$140~\kms\ at the center, likely due to beam smearing and projection effects from the highly inclined disk (i.e., inclination angle of $i \gtrsim 50\degr$). These joining effects results an “X”-shaped structure \citep{Davis17} at the second-moment map's center, reflecting steep intensity gradients across the beam \citep{Barth16b, Keppler19}. As shown in Section~\ref{sec:results}, the best-fitting dynamical models yield an intrinsic velocity dispersion of 18–22~\kms, implying that beam smearing dominates the observed linewidths across most of the \cotwo\ CND.

\begin{table}
\centering
\caption{MGE PSF model of HST/WFPC2 F814W image \label{tab:mge_psf}}
\begin{tabular}{cccc}
\hline\hline
$j$&(Light fraction)$_j$&$\sigma_j~({\rm arcsec})$&$q'_j=b_j/a_j$ \\
(1) & (2) & (3) & (4) \\
\hline
1 & 0.502 & 0.020 & 1.000\\
2 & 0.287 & 0.070 & 1.000\\
3 & 0.075 & 0.189 & 1.000\\
4 & 0.085 & 0.415 & 1.000\\
5 & 0.051 & 0.990 & 1.000\\
\hline
\end{tabular}
\noindent\parbox{\linewidth}{\textbf{Notes.} Same as Table~\ref{table:mge-ism} with column 2 lists the light fraction of each Gaussian.}
\end{table}

\begin{figure*}
\centering
\includegraphics[width=\textwidth]{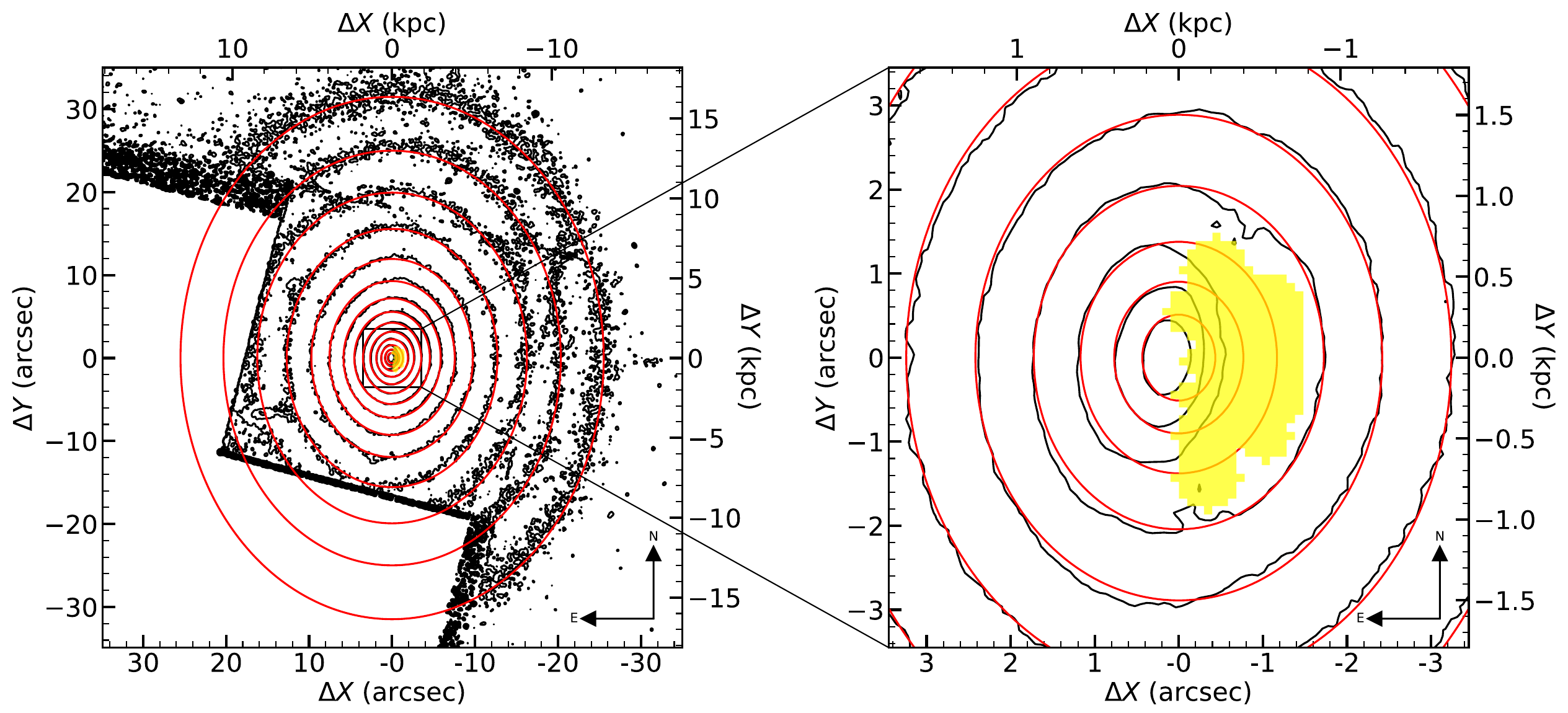}
\caption{Comparison between the HST/WFPC2 F814W image and its MGE model, shown in 2D surface brightness density over the field of $70\arcsec \times 70\arcsec$ (\textit{left}) and a central zoom ($7\arcsec \times 7\arcsec$; \textit{right}). Black contours indicate the data, and red contours show the model, demonstrating close agreement across radii and contour levels. The yellow region marks masked areas affected by bad pixels and the central dust disk.}
\label{fig:mge-fit}
\end{figure*}

\begin{table}
\centering
\caption{HST/WFPC2 F814W MGE model}    
\begin{tabular}{cccc} 
\hline\hline
$j$ &$\lg \Sigma_{\star,j}$ (\Lsun\ ${\rm pc^{-2}})$ &$\lg\sigma_j$ ($\arcsec$) &$q'_j=b_j/a_j$\\
(1) & (2) & (3) & (4)\\ 
\hline
1 & 3.937 & $-$0.396 & 0.857 \\
2 & 3.662 & $-$0.162 & 0.857 \\
3 & 3.367 &    0.095   & 0.857 \\
4 & 3.171 &    0.309   & 0.814 \\
5 & 2.928 &    0.569   & 0.814 \\
6 & 2.355 &    0.978   & 0.814 \\
7 & 0.894 &    1.385   & 0.832 \\
\hline
\end{tabular}
\parbox[t]{0.47\textwidth}{\small \textbf{Notes:} Same as Table~\ref{table:mge-ism}.} 
\label{table:mge-table}
\end{table}

\subsection{Interstellar Medium (ISM) Mass Model}\label{sec:ism}

To include the molecular gas mass ($M_{\rm gas} \approx 4.95 \times 10^8$~\Msun) in the dynamical estimation of \Mbh, we modeled its mass distribution using the Multi-Gaussian Expansion (MGE\footnote{v6.0.4: \url{https://pypi.org/project/mgefit/}}) algorithm \citep{Emsellem94}, implemented via the \textsc{Python} routine {\tt mge\_fit\_sectors\_regularized} \citep{Cappellari02}. This approach has been well tested in molecular gas modeling for dynamical BH mass measurements in NGC~3593 \citep{Nguyen22} and NGC~7052 \citep{Ngo2025a} with ALMA observations.

The zeroth-moment \cotwo\ map (panel~B of Figure~\ref{fig:moment-maps}) was converted into a molecular gas mass map and decomposed into multiple Gaussian components. This MGE decomposition was performed without deconvolving the synthesized beam, as the observational resolution had already been accounted for in the moment map creation. The resulting MGE parameters are listed in Table~\ref{table:mge-ism} and were fixed in the total mass model of NGC~4061 (i.e., with no free parameters). The good agreement between the data and the ISM MGE model is shown in Figure~\ref{fig:CO_mge}, where both are compared at identical contour levels of molecular gas-mass surface density.

Given the compactness of the continuum emission as seen in panel~A of Figure~\ref{fig:moment-maps}, much smaller than the \cotwo\ CND, and the negligible dust mass inferred from the HST optical image, we excluded the dust component from the total mass model.

\subsection{Integrated Spectrum \& Position-Velocity Diagram}\label{sec:pvd}

Panel~E of Figure~\ref{fig:moment-maps} shows the integrated \cotwo\ spectrum of NGC~4061, extracted from a $6\arcsec \times 6\arcsec$ (3.12~$\times$~3.12~kpc$^2$) aperture enclosing all line emission. The spectrum exhibits a characteristic double-horn profile, indicative of a spatially resolved rotating disk. The nearly symmetric redshifted and blueshifted sides suggest a well-settled, regularly rotating CND.

Panel~F shows the major-axis position–velocity diagram (PVD) extracted along the kinematic major axis of the CND (i.e., along the best-fitting position angle determined in Section~\ref{sec:results}; $\Gamma \approx 175\degr$). The PVD was generated using a 2-pixel-wide ($0\farcs158$) pseudo-slit, applying a spatial Gaussian filter with a FWHM equal to the synthesized beam and selecting pixels above $0.5\sigma_{\rm RMS}$ in the unsmoothed data cube. The diagram reveals a mild central rise in the intensity-weighted mean LOS velocity within the inner $\approx$0\farcs2. The CND also traces an extended  ($\approx$2\arcsec) \cotwo\ kinematics, providing tighter constraints on \ml\ and, consequently, on \Mbh\ in our dynamical models.

\section{Galaxy Mass Model}\label{sec:stellar-mass}

\subsection{HST Imaging and Photometric Model}\label{sec:smith21}

We used HST/WFPC2 F814W and F555W imaging of NGC~4061 (Program ID:~9106; PI:~Douglas Richstone), obtained on May~26,~2001, with a total exposure time from four individual exposures of 1600~s for both F814W and F555W. The data were retrieved from the Hubble Legacy Archive (HLA\footnote{\url{https://hla.stsci.edu/}}). In our fiducial analysis, we construct the stellar mass model using the HST/F814W image, and we use the F555W image to evaluate the systematic uncertainty associated with adopting an alternative photometric band in our SMBH mass measurement (see Section~\ref{sec:f555w_mge}).

We estimated the sky backgrounds for these two images by taking the median value from several $20 \times 20$~pixel$^2$ boxes located in source-free regions beyond $40\arcsec$ from the galaxy center and subtracting this median from the entire images to produce sky-subtracted frames.

To ensure accurate photometric modelings, we generated the HST/WFPC2 point spread functions (PSFs) for the F814W and F555W filters using the \textsc{TinyTim}\footnote{\url{https://github.com/spacetelescope/tinytim/releases/tag/7.5}} package \citep{Krist11}. Model PSFs of each filter were created for each of the four exposures based on the instrument configuration, chip position, and dither offsets, matching the original four-point box dither pattern. For each filter, the individual PSFs were convolved with the appropriate charge-diffusion kernel to account for CCD electron leakage. The four PSFs for each filter were then combined and resampled to a final pixel scale of $0\farcs1$ using {\tt Drizzlepac}/\textsc{AstroDrizzle}\footnote{\url{https://www.stsci.edu/scientific-community/software/drizzlepac}} \citep{Avila12}.

A mask excluding the central dust lane, bad/hot pixels, and foreground stars was created for F814W (and similar to F555W) using a point-source catalog generated with \textsc{SExtractor}\footnote{\url{https://www.astromatic.net/software/sextractor/}} \citep{Bertin96}.

We derived the stellar light distribution from the sky-subtracted, masked F814W image using the MGE method described in Section~\ref{sec:ism}. An alternative model based on the masked F555W image is presented in Section~\ref{sec:f555w_mge}. For the F814W photometric calibration, we adopted a zero point of 24.204~mag, computed from the {\tt PHOTFLAM} and {\tt PHOTPLAM} keyword in the image header following the procedures outlined in the WFPC2 Data Handbook.\footnote{\url{https://www.stsci.edu/files/live/sites/www/files/home/hst/instrumentation/legacy/wfpc2/_documents/wfpc2_dhb.pdf}} We further assumed an $I$-band solar absolute magnitude of 4.52~mag \citep{Willmer18}, with all magnitudes expressed in the AB system. The fit was deconvolved using the HST/WFPC2 F814W \textsc{TinyTim} PSF. The PSF was first decomposed into an MGE representation with the {\tt mge\_fit\_sectors} routine \citep{Cappellari02} and is provided in Section~\ref{tab:mge_psf}. This MGE PSF was then used as input for the final MGE fit of the F814W image with {\tt mge\_fit\_sectors\_regularized}, yielding a set of 2D Gaussian components convolved with the PSF MGE. The resulting deconvolved MGE can be analytically deprojected into a three-dimensional (3D) axisymmetric light distribution for an assumed inclination. The final MGE parameters are listed in Table~\ref{table:mge-table} and compared with the F814W image in Figure~\ref{fig:mge-fit}.

\subsection{Galaxy Mass Model}\label{sec:galaxy-mass}

We converted the deconvolved light-MGE into a stellar mass model for NGC~4061 by adopting a constant mass-to-light ratio parameter (from the best-fit value in Section~\ref{sec:results}) and neglecting the dark matter contribution in the central region \citep{Cappellari13a}, given the compact extent of the \cotwo\ disk. The total mass model for NGC~4061 therefore consists of three components: a central point mass representing the SMBH, the stellar mass, and the ISM mass. This combined model is used to compute the circular velocity curve for the \cotwo\ CND arising from the gravitational potential of these components.

\begin{figure}
    \centering
    \includegraphics[width=\columnwidth]{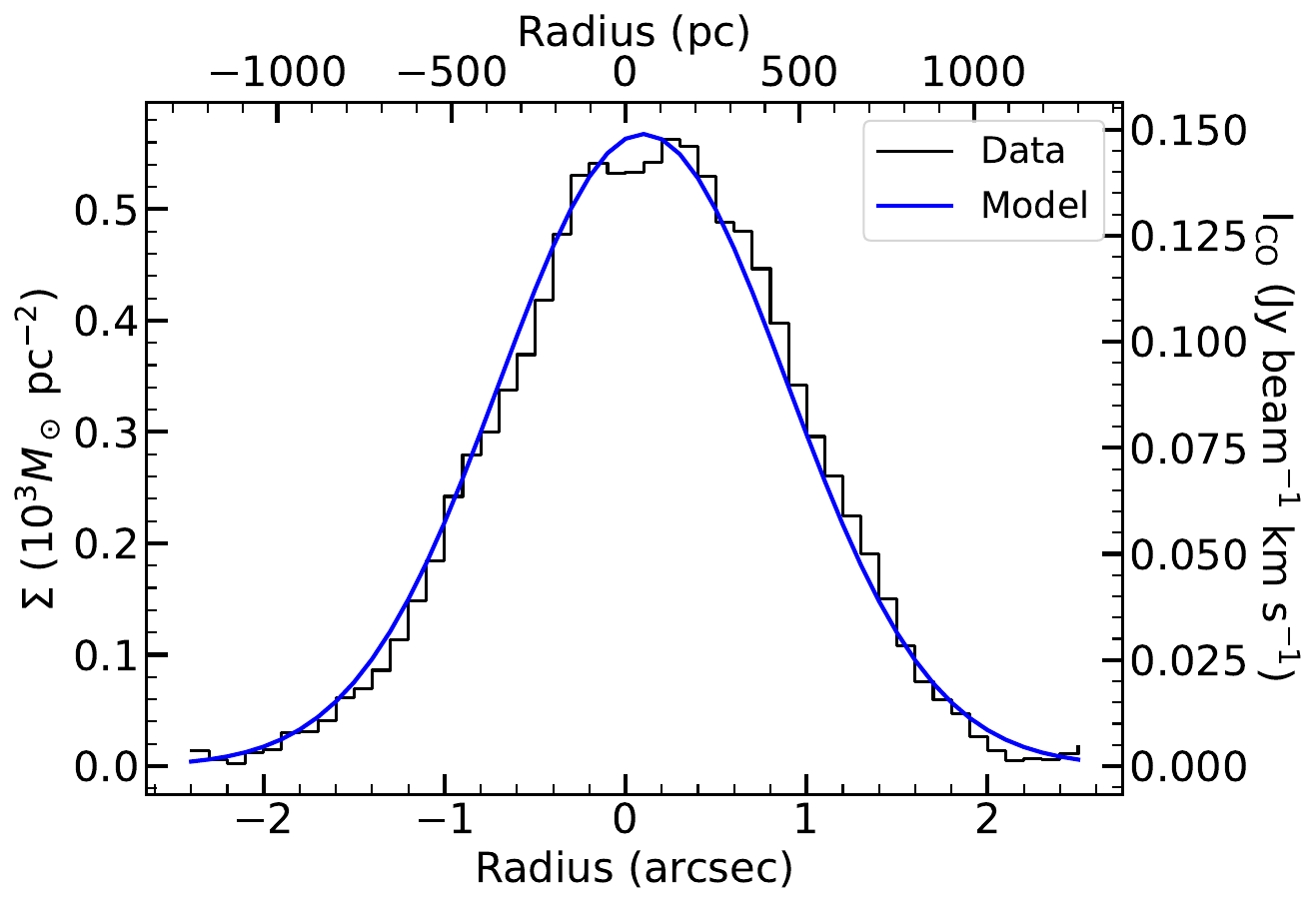}
    \caption{The \cotwo-CND surface brightness distribution of NGC~4061 is best described by a simple Gaussian model. The ALMA data are plotted in black, and the model is overlaid in blue.}
    \label{fig:axisymmetric_function}
\end{figure}

\section{Dynamical Modelling}\label{sec:dynamical-models}

\subsection{\kinms\ Tool} \label{sec:kinms}

We modeled the ALMA \cotwo-CND kinematics of NGC~4061 for its central \Mbh\ using the \textsc{Python} implementation of the KINematic Molecular Simulation tool \citep[\kinms\footnote{\url{https://github.com/TimothyADavis/KinMSpy}};][]{Davis13Nature}, which was widely applied in the WISDOM (mm-Wave Interferometric Survey of Dark Object Masses) project \citep[e.g.,][]{Davis17, Onishi17, Ruffa23} and the Measuring Black Holes in Milky Way mass galaxies (MBHBM$_\star$) project \citep{Nguyen20, Nguyen22}.

\begin{figure*}
\centering
    \includegraphics[width=\textwidth]{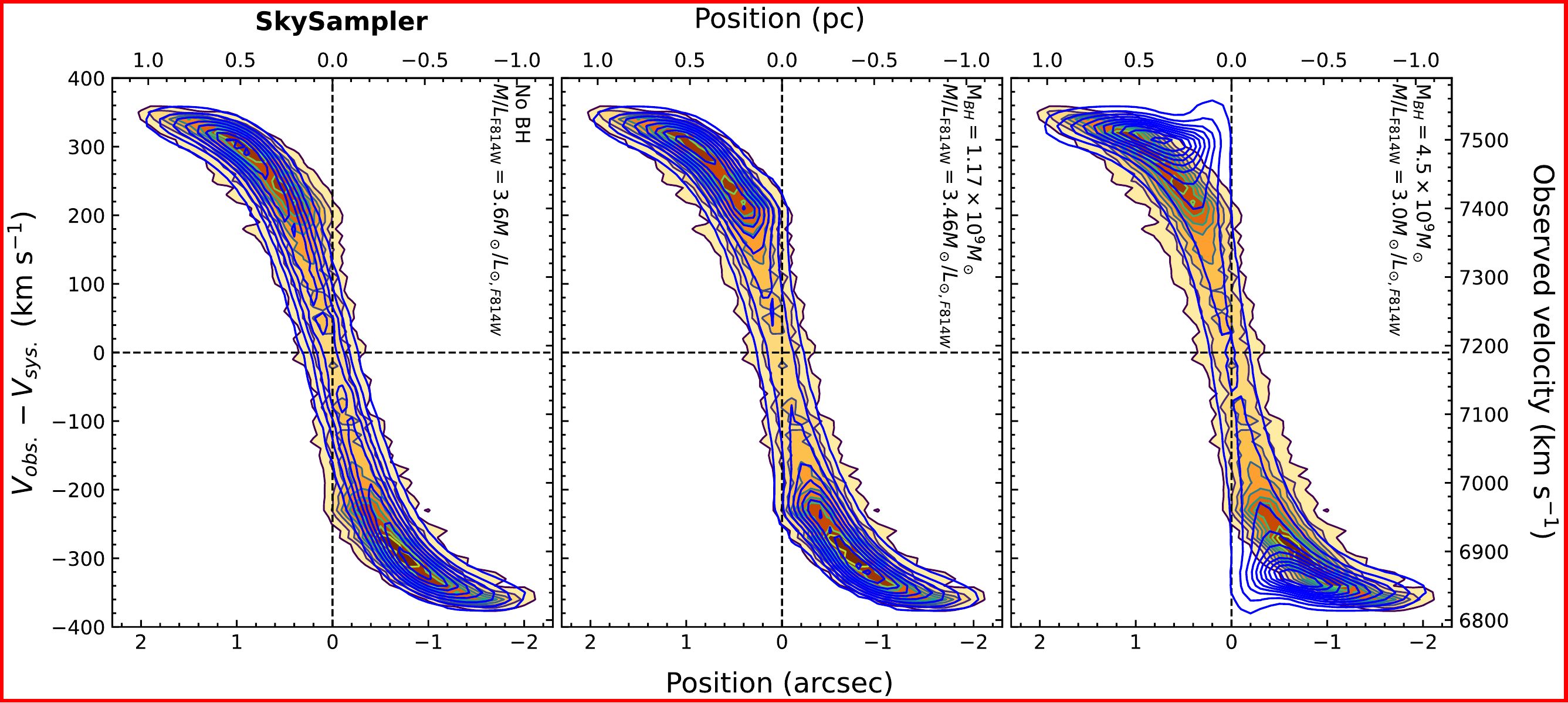}
    \includegraphics[width=\textwidth]{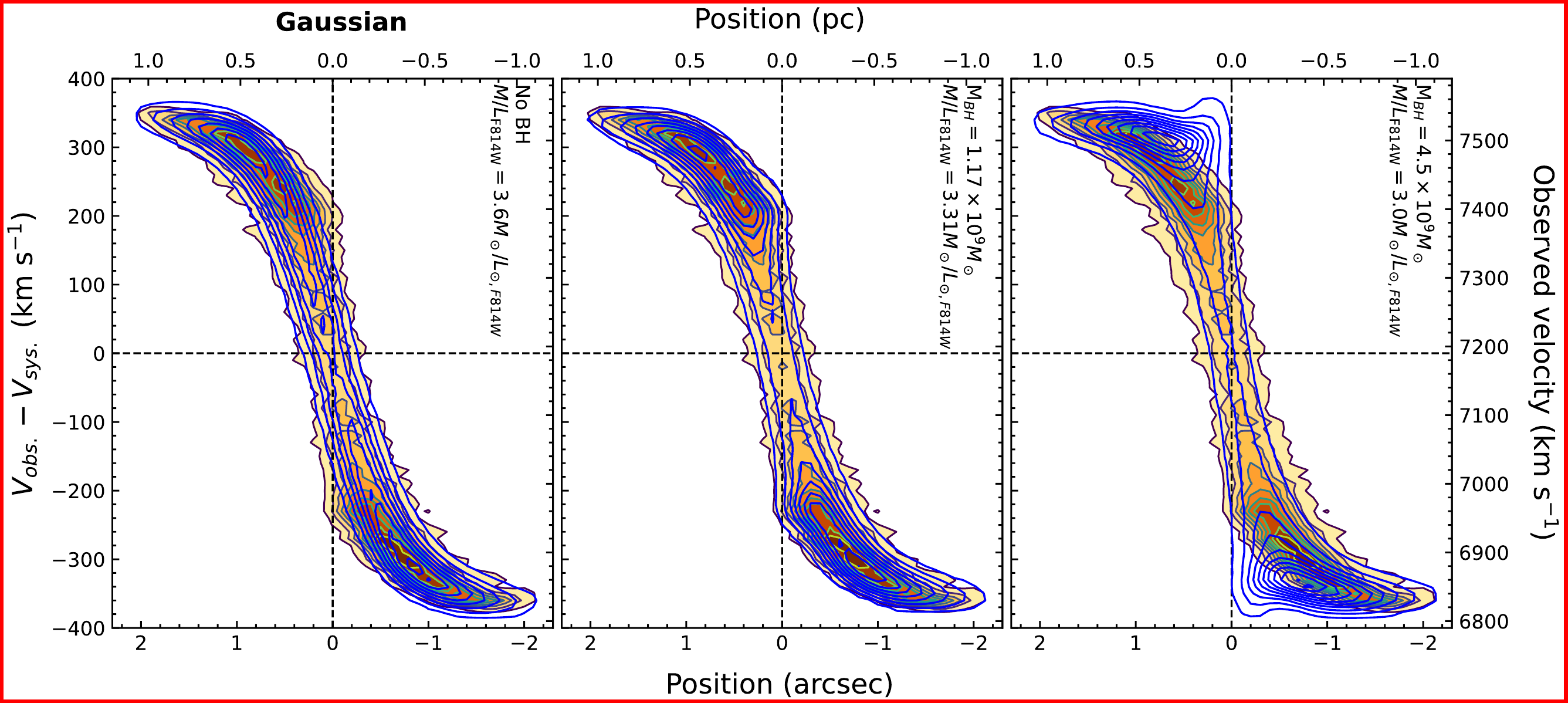}
    \caption{{\it Upper panels:} PVDs comparing the ALMA \cotwo\ observations (orange-filled contours) with \kinms\ models (blue contours) constructed using the \skysampler\ gas distribution. The PVDs are extracted along the galaxy’s major axis at a position angle of $\Gamma = 175^\circ$ for three SMBH masses: no black hole (\textit{left }), the best-fitting SMBH (\textit{middle}), and an overmassive SMBH (\textit{right}). The SMBH mass and \ml$_{\rm F814W}$ are indicated in the top-right of each panel. Black dashed lines mark the dynamical center, defined by the peak of the 1.3~mm continuum (Section~\ref{sec:cont}), with the intersection corresponding to the systemic velocity ($v_{\rm sys}=7190$~\kms) shown on the velocity axis. {\it Lower panels:} Same as above, but for \kinms\ models assuming a Gaussian gas surface brightness distribution.}
    \label{fig:compare-pvd}
\end{figure*}

\begin{figure*}
 \centering
    \includegraphics[width=\textwidth]{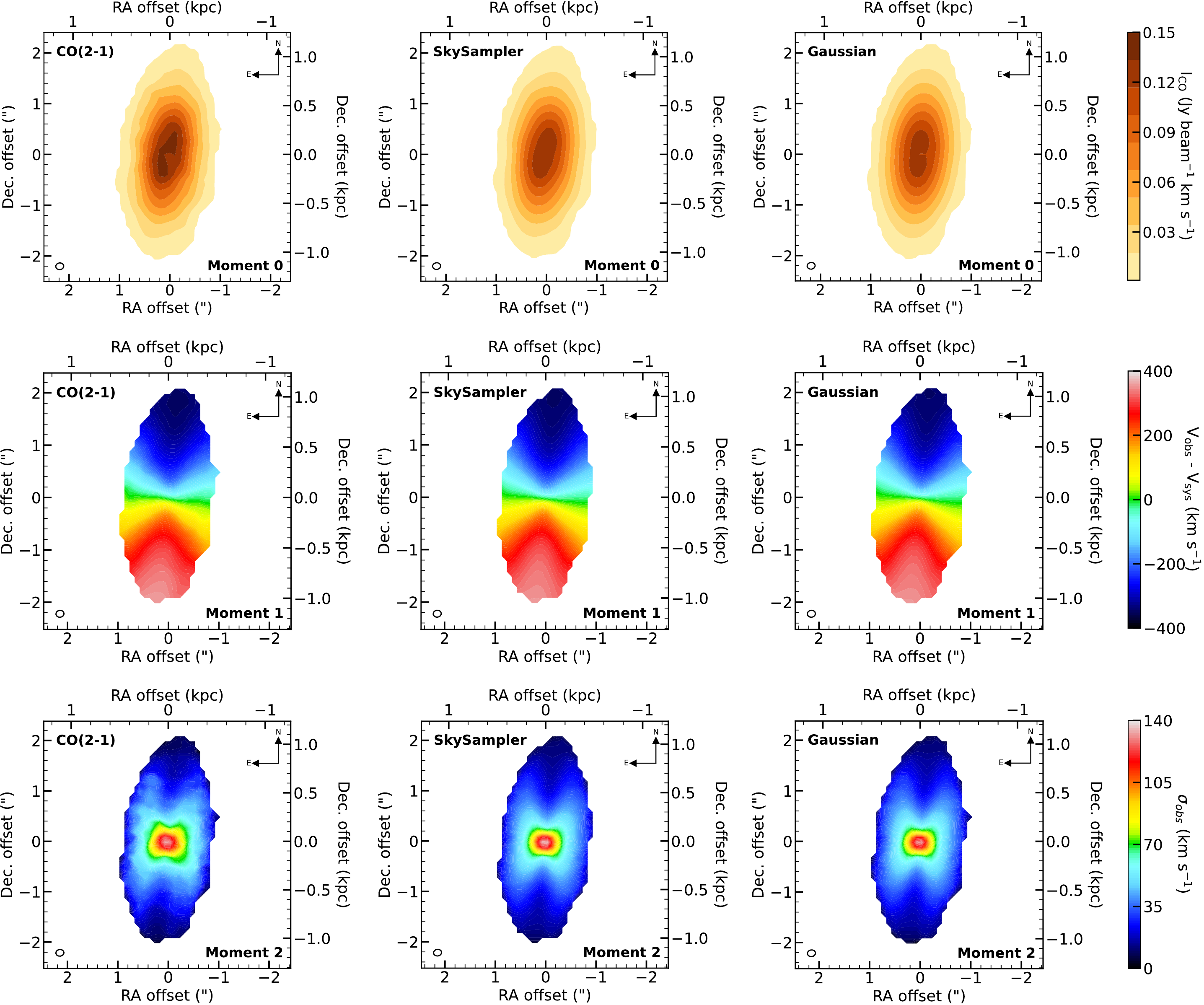}
    \caption{Comparison of the \cotwo\ moment maps from our ALMA observations (\textit{left}), the best-fitting \kinms\ model using the \skysampler\ tool to describe the gas distribution (\textit{middle}), and the best-fitting KinMS model assuming a Gaussian gas distribution (\textit{right}). All panels use identical color scales for direct comparison, demonstrating the strong consistency between the observed data and both models.}
    \label{fig:data-model-momentmaps}
\end{figure*}

KinMS generates a mock data cube by simulating the gas distribution (Section~\ref{sec:gas-dynamics}) and kinematics while incorporating observational effects such as beam smearing, spatial and velocity binning, and LOS projection. The simulated cube is compared directly with the observed cube to determine the best-fitting parameters and their uncertainties using a Markov chain Monte Carlo (MCMC) $\chi^2$ minimization routine with Bayesian priors (Section~\ref{sec:bayesian-infer}). In constructing the KinMS model, we assumed the \cotwo\ gas follows circular orbits around the galactic center under the combined gravitational potentials of the SMBH, stellar, and gas components. The circular velocity as a function of radius was computed using the {\tt mge\_vcirc} routine from the Jeans Anisotropic Modeling framework \citep[\textsc{JAM}\footnote{v7.2.4: \url{https://pypi.org/project/jampy/}};][]{Cappellari08}.

Our adopted \kinms\ models fit the observed data by optimizing nine free parameters. The first two are the kinematic center coordinates ($x_{\rm cen}$ and $y_{\rm cen}$), which specify the SMBH position relative to the data cube phase center or the continuum peak identified in Section~\ref{sec:cont}. This assumption is justified, as any offset between the kinematic and photometric centers is typically much smaller than the synthesized beam. The third parameter is the systemic velocity of the gas disc ($v_{\rm sys}$), or equivalently the velocity offset ($v_{\rm off}$) if $v_{\rm sys}$ has already been subtracted. The fourth is the integrated intensity scaling factor ($f$, Section~\ref{sec:gas-dynamics}), describing the normalization of the gas distribution.   The next three parameters characterize the CND morphology: inclination ($i$), position angle ($\Gamma$), and the intrinsic turbulent velocity dispersion of the gas ($\sigma_{\rm gas}$). The final two parameters are the SMBH mass (\Mbh) and the stellar mass-to-light ratio in the F814W band (\ml$_{\rm F814W}$). These nine free parameters are summarized in Table~\ref{tab:mcmc-results}.

\begin{table*}
\caption{Best-fitting \kinms\ parameters and their uncertainties}
\centering
\begin{tabular}{lcccccr}
\hline \hline
Model parameters&Search range& Initial guesses&Best-fit values&1$\sigma$ (16--84\%)&3$\sigma$ (0.14--99.86\%)\\ 
(1)             & (2)        &      (3)       &      (4)      &           (5)      & (6)\\ 
\hline\hline
\multicolumn{6}{c}{\skysampler\ ($\chi^2_{\rm red, min} \approx 0.80$)}\\ 
\hline
\underline{Mass model:}        & ~ & ~ & ~ & ~ & ~ \\  
$\lg(M_{\rm BH}$/\Msun)        & 7 $\rightarrow$ 11  & 9 & 9.07 & +0.03,$-$0.04 & +0.10,$-$0.11 \\ 
$M/L_{\rm F814W}$ (\Msun/\Lsun)& 0 $\rightarrow$ 5   & 3.5 & 3.46 & +0.07,$-$0.06 & +0.20,$-$0.17 \\ [1mm]
\underline{\cotwo\ CND:}       & ~ & ~ & ~ & ~ & ~ \\ 
$f$ (Jy \kms)                  & 1  $\rightarrow$ 50  & 25  & 25.78  & +0.71,$-$0.70 & +2.06,$-$2.00 \\ 
$i$ (\degr)                    & 42 $\rightarrow$ 90  & 60  & 60.41  & +1.03,$-$1.11 & +2.80,$-$3.36 \\ 
$\Gamma$ (\degr)               & 150$\rightarrow$ 200 & 175 & 174.62 & +0.50,$-$0.50 & +1.44,$-$1.47 \\ 
\siggas (\kms)                 & 0  $\rightarrow$ 50  & 20  & 13.91  & +1.98,$-$1.93 & +5.71,$-$5.29 \\ [1mm]
\underline{Nuisance:}          & ~ & ~ & ~ & ~ & ~ \\
$x_c$ (arcsec)                 & $-$0.2 $\rightarrow$ +0.2 & 0 & $-$0.03     & +0.01,$-$0.01 & +0.02,$-$0.02 \\ 
$y_c$ (arcsec)                 & $-$0.2 $\rightarrow$ +0.2 & 0 & 0.02  & +0.01,$-$0.01 & +0.02,$-$0.03 \\ 
$v_{\rm off}$ (\kms)           & $-$50  $\rightarrow$ +50  & 0 & $-$10.24 & +1.61,$-$1.63 & +4.67,$-$4.77 \\ 
\hline\hline
\multicolumn{6}{c}{Gaussian ($\chi^2_{\rm red, min} \approx 0.78$)}\\ 
\hline
\underline{Mass model:}        & ~ & ~ & ~ & ~ & ~ \\ 
$\lg(M_{\rm BH}$/\Msun)        & 7 $\rightarrow$ 11 & 9 & 9.07 & +0.04,$-$0.04 & +0.10,$-$0.12 \\ 
$M/L_{\rm F814W}$ (\Msun/\Lsun)& 0 $\rightarrow$ 5  & 3.5 & 3.31 & +0.06,$-$0.06 & +0.18,$-$0.16 \\ [1mm]
\underline{\cotwo\ CND:}       & ~ & ~ & ~ & ~ & ~ \\ 
$f$ (Jy \kms)                  & 1 $\rightarrow$ 50 & 25& 27.20& +0.84,$-$0.84 & +2.56,$-$2.40 \\ 
$i$ (\degr)                    & 42$\rightarrow$ 90 & 60& 62.11& +0.90,$-$0.94 & +2.60,$-$2.86 \\ 
$\Gamma$ (\degr)               &150$\rightarrow$ 200&175&174.36& +0.51,$-$0.51 & +1.47,$-$1.47 \\  
\siggas (\kms)                 & 0 $\rightarrow$ 50 & 20& 15.51& +2.03,$-$1.92 & +6.18,$-$5.41 \\
\hline
\end{tabular}
\parbox[t]{0.98\textwidth}{\textbf{Notes:} When modeling the gas distribution with a simple Gaussian profile in \kinms\, we fixed the nuisance parameters to their best-fit values obtained from the previous case that used the \skysampler\ tool to constrain the gas distribution.}
\label{tab:mcmc-results}
\end{table*}

\subsection{Gas Distribution} \label{sec:gas-dynamics}

Because our modeling fits the full 3D ALMA cube, a parameterized gas distribution is required and scaled by the integrated intensity factor ($f$) to match the observations. In this work, we represent the gas distribution using either the \skysampler \footnote{\url{https://github.com/Mark-D-Smith/KinMS-skySampler}} CLEAN-component model derived directly from the data cube (Section~\ref{sec:SkySampler}) or an analytic axisymmetric profile (Section~\ref{sec:axisymmetric_function}).

\subsubsection{Gas Distribution Assumed by \skysampler} \label{sec:SkySampler}

\skysampler\ constructs the molecular gas clouds in the \cotwo\ CND of NGC~4061 directly from the CLEAN components of the data cube, allowing the model to fit only the gas kinematics without assuming a specific spatial distribution. This approach introduces a single free parameter—the total flux scaling factor ($f$)—which rescales the cube to match the observed flux. Because CLEAN components omit residual emission from the deconvolution process, their total flux is slightly lower than that of the original cube; $f$ compensates for this difference and ensures flux conservation within the modeled region.

We uniformly sampled the CLEAN components with $10^6$ gas particles using the {\tt sampleClouds} routine, reproducing the observed CO surface brightness distribution after beam convolution. The particles were then deprojected from the sky plane to the intrinsic galaxy plane with {\tt transformClouds}, adopting a position angle of $\Gamma = 175\degr$ and an inclination of $i = 60\degr$.

Although Panel~D of Figure~\ref{fig:moment-maps} shows spatial variations in velocity dispersion, these are largely caused by beam smearing and projection effects in the highly inclined disc. We therefore adopted a constant \siggas\ and modeled the gas as a geometrically thin disk with zero scale height in our \kinms\ simulations.

\begin{figure*}
    \centering
    \includegraphics[width=\textwidth]{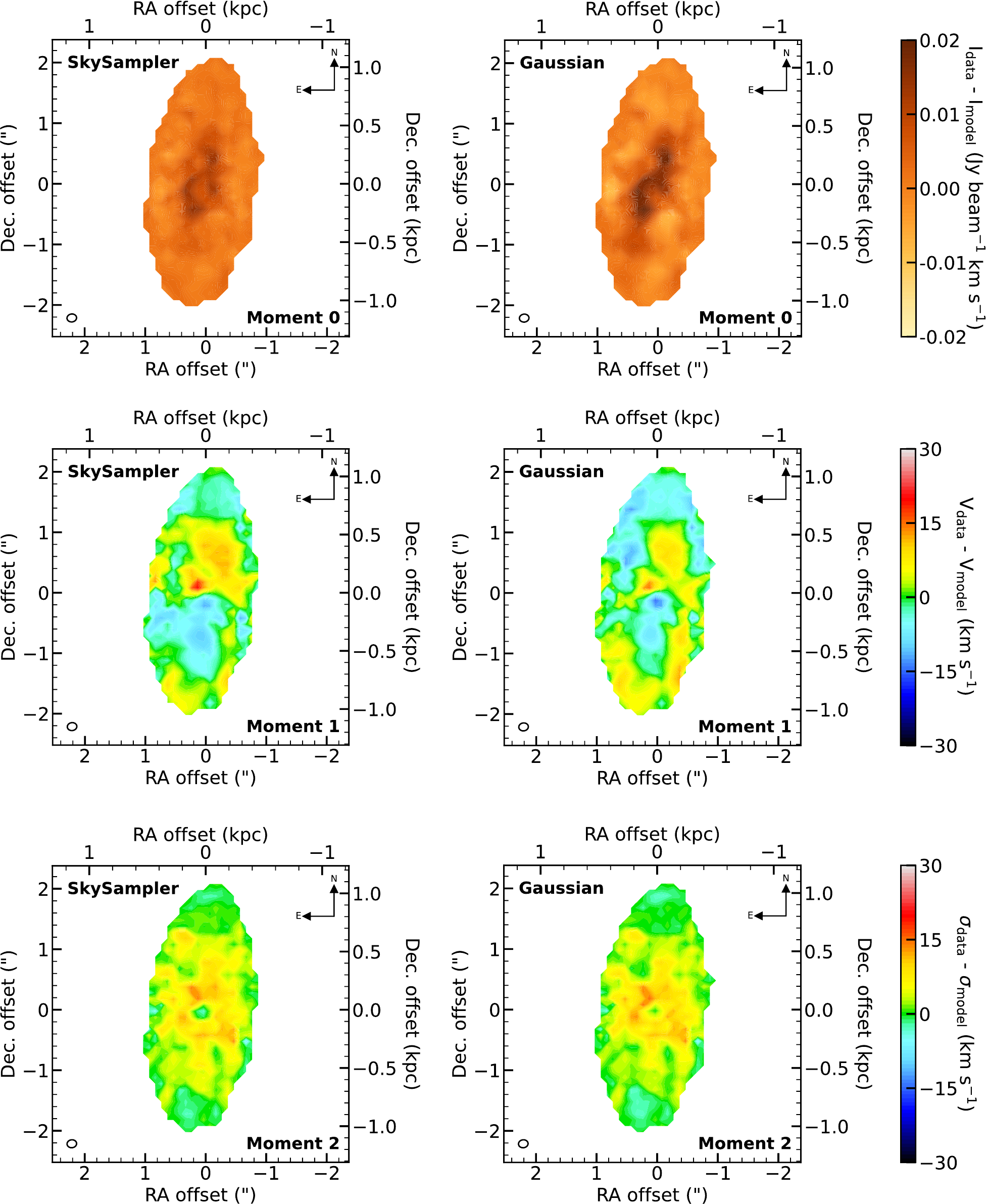}
    \caption{Moment-residual maps ({\tt data-model}), obtained by subtracting the moment maps of the best-fitting \kinms\ models from the observations. The residuals are $\lesssim$10\% for the model using \skysampler\ and $\lesssim$13\% for the Gaussian model, demonstrating good agreement between the data and both models and indicating no significant non-circular motions or warps in the \cotwo-CND of NGC~4061.}
    \label{fig:overlaid-spec-residual} 
\end{figure*}

\subsubsection{Gaussian Distribution} \label{sec:axisymmetric_function}

Given the smooth, centrally concentrated morphology of the \cotwo\ CND (Section~\ref{sec:ism}), we alternatively modeled it with a single Gaussian profile \citep[e.g.,][]{Nguyen20} without deprojection (fixed $q_j = 1$), as shown in Figure~\ref{fig:axisymmetric_function}. This Gaussian profile is centered at an offset of $\mu = 0\farcs09$ (46.8 pc) with a dispersion of $\sigma = 0\farcs79$ (410.8 pc, corresponding to ${\rm FWHM} = 1\farcs87$ or 972.4 pc); these parameters are held fixed in the corresponding \kinms\ models. In this case, only the amplitude parameter ($f$) is allowed to vary, with the same interpretation as in Section~\ref{sec:SkySampler}.         

\begin{figure*}
\hspace{-10mm}
    \includegraphics[width=1.1\textwidth]{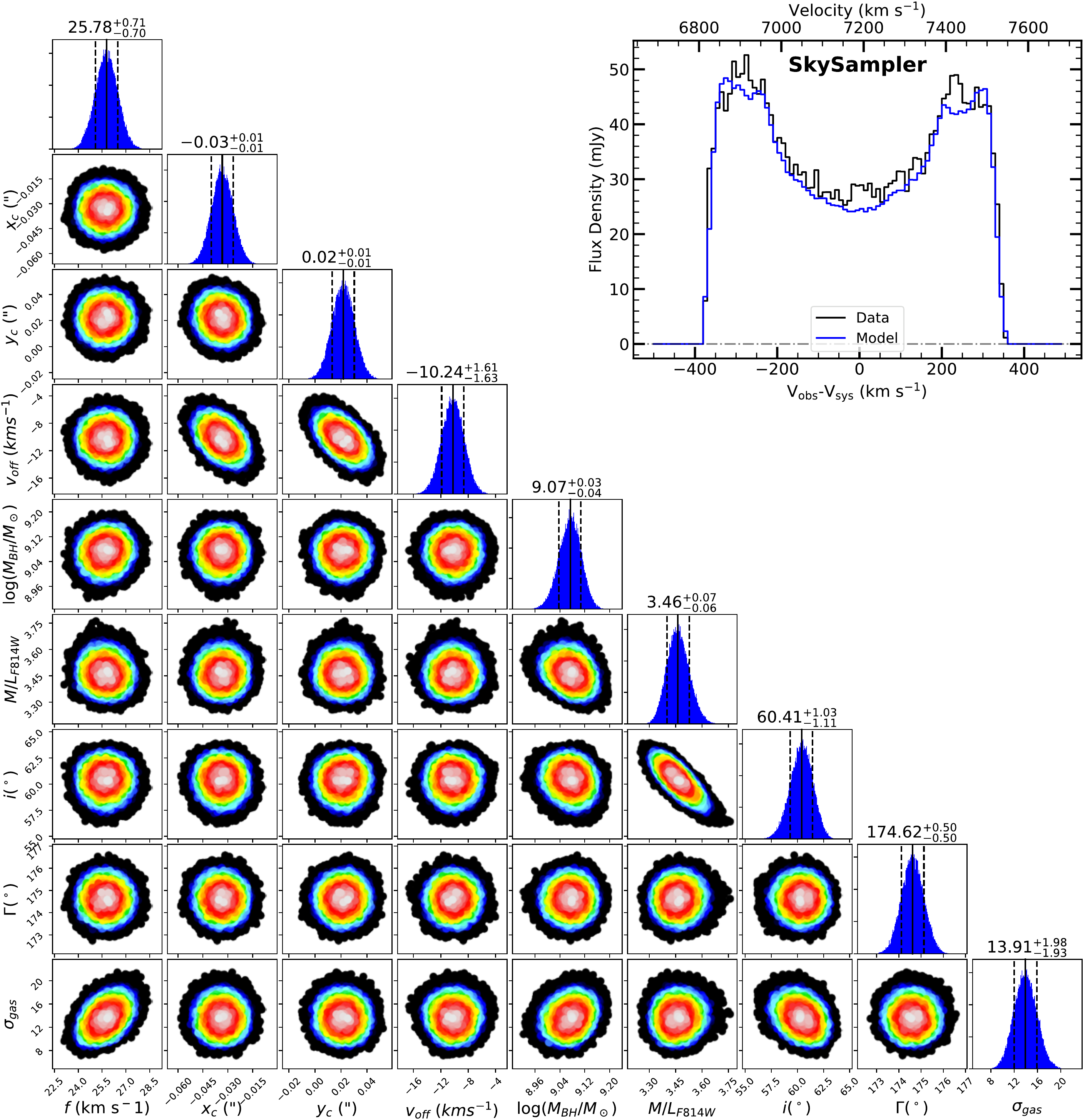}
    \caption{{\it Triangle:} Corner plot showing the posterior distributions from the \kinms\ model assuming the gas distribution derived with the \skysampler\ tool. The top panels display the 1D marginalized posterior distributions for each parameter, with vertical lines indicating the best-fit values and 1$\sigma$ uncertainties (see text for details). The lower panels show 2D projections of parameter pairs, with colors indicating CLs: white, red, green, and black correspond to the best-fit 1$\sigma$, 2$\sigma$, 3$\sigma$, and larger than 3$\sigma$ CLs, respectively. The numerical results are summarized in Table~\ref{tab:mcmc-results}. {\it Insert panel:} Comparison of the ALMA \cotwo\ integrated spectrum (panel E of Figure~\ref{fig:moment-maps}) with the same profile extracted from the best-fitting \kinms\ model, which highlights the good agreement between the data and model.} 
    \label{fig:triangle}
\end{figure*}

\subsection{Bayesian Inference and Priors} \label{sec:bayesian-infer}

We employed the adaptive Metropolis algorithm \citep{Haario01}, implemented within a Bayesian framework using the \textsc{AdaMet}\footnote{v2.0.9; \url{https://pypi.org/project/adamet/}} package \citep{Cappellari13a}, to constrain the best-fitting parameters of the \kinms\ model and estimate their statistical and formal uncertainties, which are propagated from the \cotwo-CND kinematics measured from our ALMA observations. Each MCMC chain consisted of $10^5$ iterations, with the first 20\% discarded as burn-in. The remaining samples were used to construct the posterior probability distribution functions (PDFs). Best-fit values were taken as the parameters corresponding to the maximum likelihood, and uncertainties were estimated at the 1$\sigma$ (16–84\%) and 3$\sigma$ (0.14–99.86\%) confidence levels (CLs). Because the search range of the \Mbh\ parameter spans several orders of magnitude, it was sampled logarithmically, while all other parameters were sampled linearly. Convergence and adequate sampling of the parameter space were verified through inspection of the MCMC chains, using the parameter ranges and initial guesses summarized in Section~\ref{tab:mcmc-results}.

In the Bayesian framework, the priors are proportional to the logarithm of the likelihood, $\ln{(\rm data|model)} \propto 0.5\chi^2$, where $\chi^2$ is defined as:
\begin{equation*}
\chi^2 \equiv \sum_i{\frac{({\rm data}_i - {\rm model}_i)^2}{\sigma_i^2}} = \frac{1}{\sigma_{\rm RMS}^2}\sum_i{({\rm data}_i - {\rm model}_i)^2},
\label{eq:chi2}
\end{equation*}
where $\sigma_{\rm RMS}$ is defined by the mask in Section~\ref{sec:momentmaps} and were assumed as a constant $\sigma$ for all pixels. When computing $\chi^2$, we rescaled the data cube uncertainties by a factor of $(2N)^{0.25}\approx20$, where $N=78,100$ is the number of pixels with detected emission. This scaling yields more realistic fit uncertainties by accounting for underestimated systematic effects commonly encountered in Bayesian analyses of large datasets such as ALMA \citep{Ngo2025a}, VLT/MUSE \citep{Thater22, Thater23}, and JWST/NIRSpec \citep{Nguyen2025c, Nguyen2026a} observations. These effects arise from correlated noise between adjacent pixels, a result of the synthesized beam size inherent to interferometric data cube—an effect known as ``noise covariance'' \citep{Onishi17, North19, Nguyen20}. This approach was first proposed by \citet{vandenBosch09}, later refined by \citet{Mitzkus17}, and subsequently adopted in multiple WISDOM \citep[e.g.,][]{North19, Smith19} and MBHBM$\star$ \citep[e.g.,][]{Nguyen19conf} studies.

\begin{figure*}
    \centering
    \includegraphics[width=\textwidth]{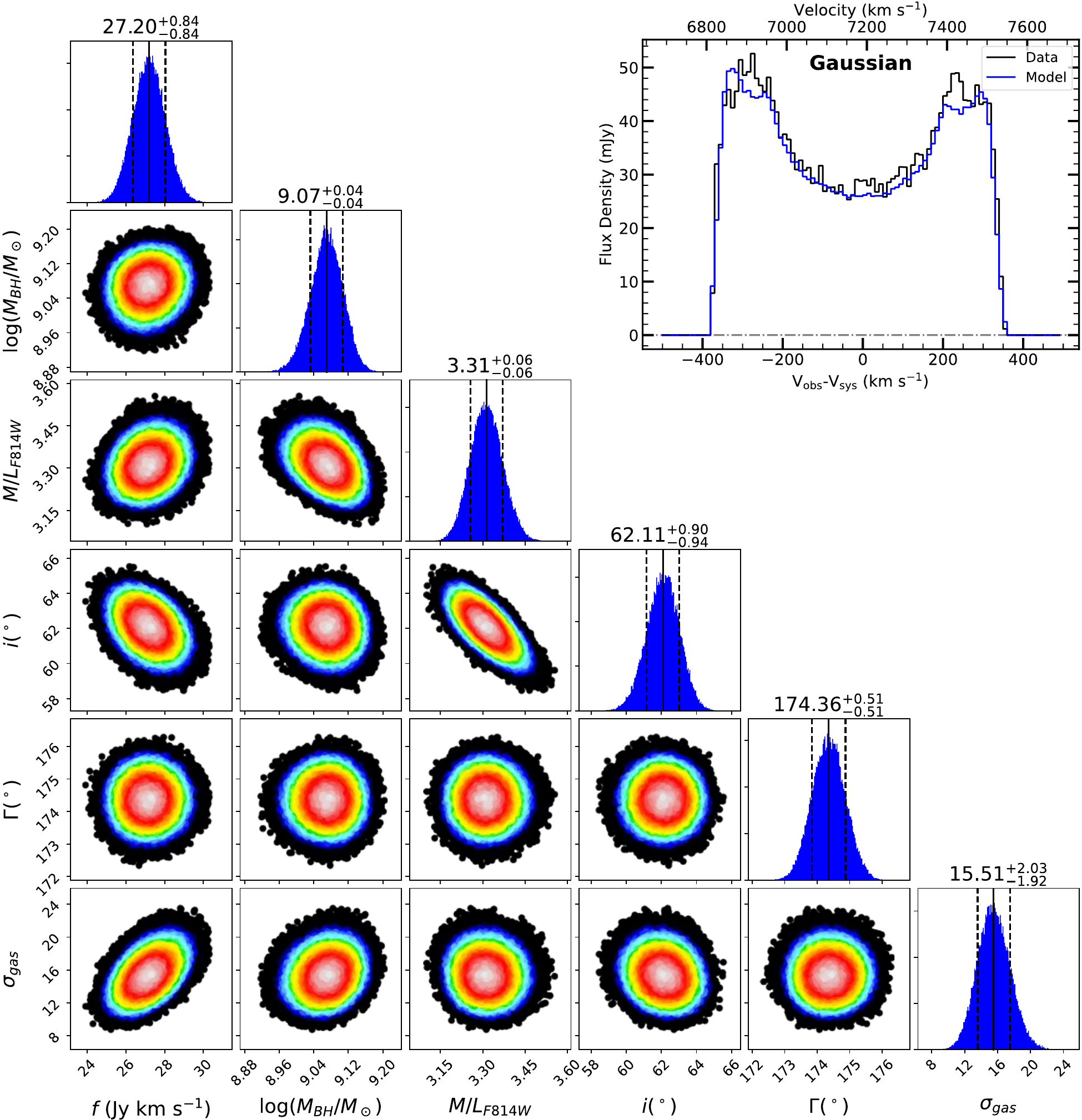}
    \caption{Same as Figure~\ref{fig:triangle} but the \kinms\ model was assumed to be a Gaussian for gas surface brightness distribution.}  
     \label{fig:triangle1}
\end{figure*}

\subsection{Results}\label{sec:results}

Given the synthesized beam size of our combined ALMA \cotwo\ observations is approximately equal to the expected SOI of the tentative SMBH, the comparison between the observed molecular gas kinematics and our \kinms\ models reveals clear evidence for a central SMBH, as indicated by a modest rise in rotation velocity toward the nucleus at radii smaller than $0\farcs4$ (see Figure~\ref{fig:compare-pvd}). As summarized in Table~\ref{tab:mcmc-results}, the best-fitting \kinms\ model using the \skysampler\ tool yields \Mbh\ $=(1.17^{+0.08}_{-0.10})\times10^9$~\Msun\ and $M/L_{\rm F814W}=3.46^{+0.07}_{-0.06}$~(\Msun/\Lsun). A comparable model employing a Gaussian disk provides \Mbh\ $=(1.17^{+0.11}_{-0.10})\times10^9$~\Msun\ and $M/L_{\rm F814W}=3.31\pm0.06$~(\Msun/\Lsun). The \skysampler\ model achieves a minimum $\chi^2_{\rm min}=62,472$, corresponding to a reduced $\chi^2_{\rm red,min}=0.80$, while the Gaussian model yields $\chi^2_{\rm min}=60{,}913$ and $\chi^2_{\rm red,min}=0.78$. All quoted uncertainties represent $1\sigma$ confidence intervals (CLs) unless otherwise stated; both $1\sigma$ and $3\sigma$ CLs are reported in Table~\ref{tab:mcmc-results} and Table~\ref{tab:mcmc-results_mlvary}.

Both best-fitting models reproduce the ALMA \cotwo\ data remarkably well across all spatial positions of the CND. This agreement is evident in the middle panels of Figure~\ref{fig:compare-pvd} and in Figure~\ref{fig:data-model-momentmaps}, which compare the observed and modeled PVDs) along the major axis and the corresponding moment maps directly and respectively.

In each modeling approach, whether using \skysampler\ or a Gaussian surface brightness distribution, Figure~\ref{fig:compare-pvd} also includes two comparative \kinms\ models. The first assumes no SMBH (\Mbh\ $=0$~\Msun) with $M/L_{\rm F814W}=3.6$~(\Msun/\Lsun), which reproduces the extended CND kinematics beyond $r>0\farcs4$ but fails to match the slightly central rise in rotation velocity ($\approx$80~\kms). This no-SMBH model also modestly overpredicts ($\approx$35~\kms) the CDN's rotation at radii beyond 1\arcsec. The second adopts a larger fixed SMBH mass of \Mbh\ $=4.5\times10^9$~\Msun\ with $M/L_{\rm F814W}=3.0$~(\Msun/\Lsun); it similarly fits the outer kinematics but overpredicts the central velocity increase at small radii. In all these alternative models, only $M/L_{\rm F814W}$ was allowed to vary, while \Mbh\ was held fixed at the specified values and all other parameters were kept at their best-fit values listed in Table~\ref{tab:mcmc-results}.

Figure~\ref{fig:overlaid-spec-residual} further evaluates the consistency between our ALMA observations and the best-fitting \kinms\ models by examining the residual maps of all three moments to assess which model provides the better representation visually. The residuals show no discernible structures in either the integrated intensity (top panels) or the intensity-weighted mean LOS velocity dispersion maps (bottom panels). Any signatures of non-circular motions (e.g., inflows or outflows) within the \cotwo\ CND are negligible, as indicated by the intensity-weighted mean LOS velocity residuals ($V_{\rm residual}=V_{\rm data} - V_{\rm model}$; middle panels). The best-fitting \kinms\ model using the \skysampler\ tool yields residual velocities of $|V_{\rm residual}| \lesssim 15$~\kms\ ($\lesssim$4\%) across the CND, comparable to the spectral channel width of the reduced ALMA cube ($\approx$10~\kms), implying an absence of significant non-circular motions. The Gaussian-based model produces slightly smaller residuals of $|V_{\rm residual}| \lesssim 10$~\kms\ ($\lesssim$3\%), likely due to its assumption of a smooth gas distribution, which only approximates the observed morphology. In contrast, the \skysampler-based model incorporates the actual spatial gas distribution from the data cube, substantially reducing discrepancies in the intensity-weighted mean LOS velocity field. In this work, we adopt the average of both the \skysampler-based and the Gaussian-based best-fitting model as our fiducial result.

Figure~\ref{fig:triangle} (for the \skysampler-based \kinms\ models) and Figure~\ref{fig:triangle1} (for the Gaussian-based \kinms\ models) display the 2D posterior distributions for each pair of free parameters, marginalizing over the others, with colors indicating likelihood levels. White corresponds to the maximum likelihood within the 1$\sigma$ CL, while blue denotes the 3$\sigma$ CL. The 1D histograms along the diagonal show the marginalized distributions for each parameter, where the thick black vertical lines indicate the best-fit values and the dashed lines mark their 1$\sigma$ uncertainties. All distributions exhibit Gaussian-like shapes, demonstrating that the MCMC optimization of our \kinms\ models achieved robust convergence.

In both figures, we further assess the consistency between the observed \cotwo\ emission and the best-fitting \kinms\ models by comparing their integrated spectra in the inset panels. The best-fit models not only reproduce the overall kinematic structure but also recover the symmetric double-horn profile of the total integrated \cotwo\ spectrum across the CND. These results confirm that the best-fitting models provide an accurate representation of the observed molecular gas kinematics.

\begin{figure}
    \centering
    \includegraphics[width=\columnwidth]{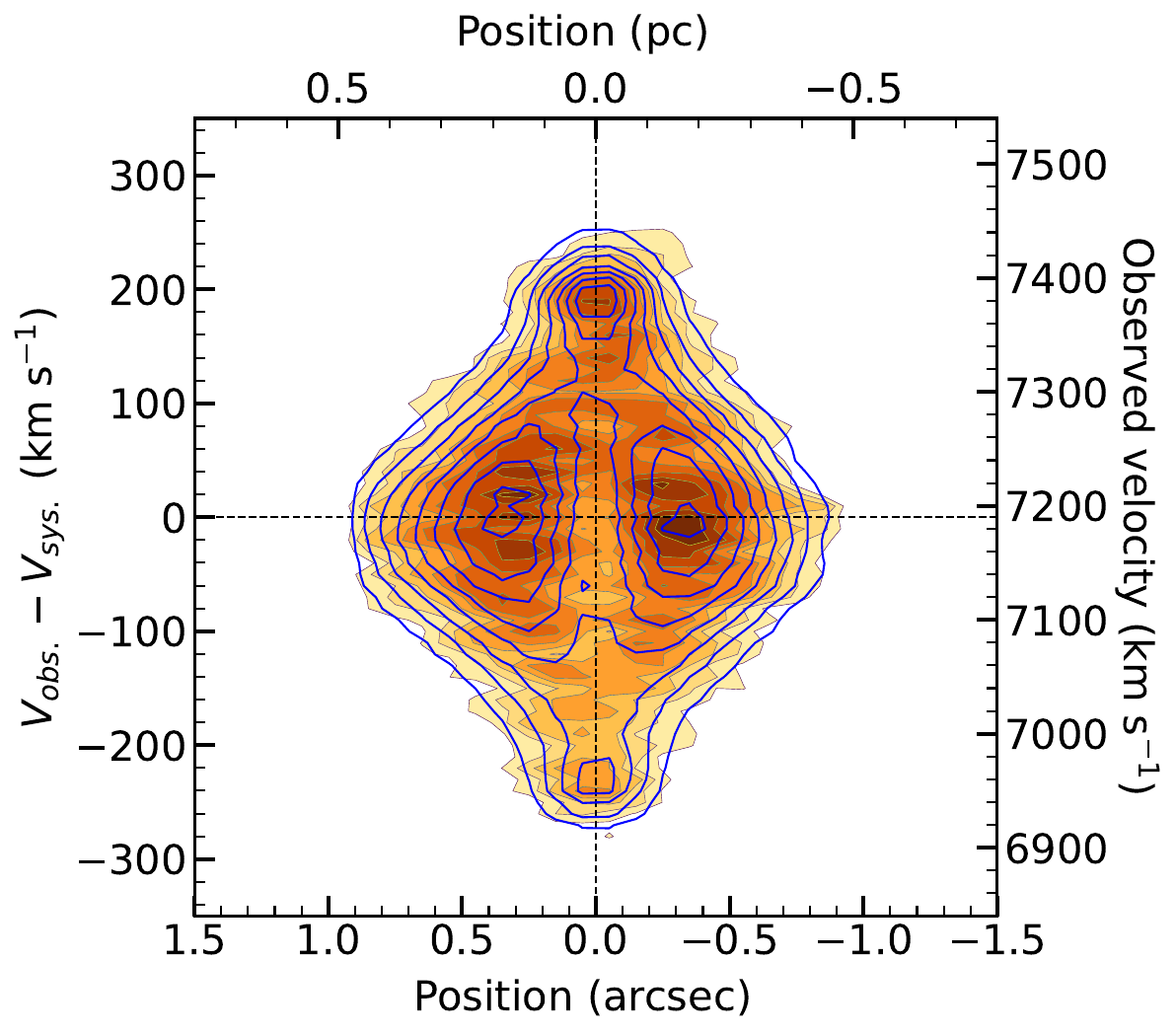}
    \caption{PVD extracted along the minor axis of the \cotwo\ CND, oriented at a position angle of $\Gamma = 175^{\circ} + 90^{\circ}$, with a systemic velocity of $v_{\rm sys} = 7190$ km s$^{-1}$. The best-fitting \skysampler-based (blue) KinMS model is overplotted at the same contour levels.}
    \label{fig:PVD_minoraxis}
\end{figure}

We also verified the absence of significant non-circular motions or kinematic warps (i.e., variations in position angle, $\Gamma$, that would twist the isovelocity contours along the CND minor axis). Such effects could bias the dynamical modeling and \Mbh\ estimation. As shown in Figure~\ref{fig:PVD_minoraxis}, the minor-axis PVD of NGC~4061, extracted along $\Gamma~(\approx175\degr) + 90\degr$, exhibits symmetry across all four “forbidden quadrants,” a feature also well reproduced by the best-fitting \skysampler-based \kinms\ model, confirming the overall regular rotation of the \cotwo\ CND.  

While the other model parameters are well constrained, a clear anti-correlation between \Mbh\ and $M/L_{\rm F814W}$ is evident (see both Figure~\ref{fig:triangle} and Figure~\ref{fig:triangle1}). This degeneracy is commonly observed in spatially resolved dynamical modeling and arises because both stars and BH gravitational potentials affect the central circular motions of the \cotwo\ CND. 

Given that the angular resolution of our combined ALMA data matches the SMBH sphere of influence (SOI; the region surrounding a BH within which its gravitational influence dominates that of the surrounding stellar mass; equivalently, the radius ($r_{\rm soi}$) where the enclosed stellar mass equals \Mbh. It can also be calculated through the stellar velocity dispersion ($\sigma$) as $r_{\rm soi}=GM_{\rm BH}/\sigma^2$.), implying that the SMBH SOI is resolved and explaining the observed \Mbh--$M/L_{\rm F814W}$ anti-correlation in the posterior distributions.

Additional covariances are also present among other parameters:
\begin{description}
    \item[$y_{\rm cen}$ and $v_{\rm off}$] resulting from the large beam size when the kinematic center is tied to the peak of the spatially resolved continuum emission (Section~\ref{sec:cont}). Higher angular resolution observation will tighter constrain the CND disk center and suppress this degeneracy. 
    \item[$i$ and $M/L_{\rm F814W}$] which emerges as the CND transitions from more edge-on to more face-on geometries \citep{Smith19}; the moderate inclination of the \cotwo\ CND in NGC~4061 ($i \approx 59^{\circ}$) reflects this behavior.
    \item[$f$ and $\sigma_{\rm gas}$] since $f$ tends to normalize to the adopted surface brightness distribution \citep{Smith19, Nguyen20}.
\end{description}    
Such parameter covariances are typical in simultaneous dynamical modeling of molecular gas disks. 

Figure~\ref{fig:NGC4061_cumulative_mass} presents the enclosed mass profiles of the SMBH, stellar component, and molecular gas within $2\arcsec$ of the \cotwo\ CND. Although the molecular gas contributes nearly two orders of magnitude less mass than the stars over this region, the black hole dominates the gravitational potential inside $r \lesssim 0\farcs2$. Moreover, the synthesized beam of the combined ALMA data cube ($\theta_{\rm beam} \approx 0\farcs13$) closely matches the estimated SOI radius of the SMBH ($r_{\rm soi} \approx 0\farcs12$), calculated using our best-fit black hole mass and an assumed stellar velocity dispersion of $\sigma \approx 290$~\kms\ \citep{Pinkney2005}. This close correspondence between the observational angular resolution and the SMBH SOI provides strong support for the reliability of our \Mbh\ measurement in NGC~4061.

\subsection{Uncertainties} \label{sec:uncertainties}

Several additional sources of uncertainty are inherent to dynamical modeling, including (i) the adopted distance to NGC~4061, (ii) the assumption of a thick disk (implemented via the $z$-coordinate perpendicular to the disk plane), (iii) the turbulent velocity dispersion of the gas, (iv) the disk inclination, (v) possible radial variations in the mass-to-light ratio, (vi) the construction of the stellar mass model without dust masking, and (vii) the use of an alternative photometric band. We discuss each of these effects in the following subsections.

\subsubsection{Distances}\label{sec:uncer_dist}

Since the inferred \Mbh\ scales linearly with the assumed distance ($M_{\rm BH} \propto D$), the wide range of distance estimates for NGC~4061 (68–112 Mpc; NED)—primarily based on optical and near-infrared light curves of the Type Ia supernova SN~2008bf—constitutes the dominant source of uncertainty in our \Mbh\ measurement. We adopt a distance of 107.2 Mpc from the MASSIVE survey, derived from the galaxy’s redshift \citep{Ma14}. The spread in published distances introduces a systematic uncertainty of $\approx$37\% in \Mbh.

\begin{table}
\caption{Best-fitting \ml\ models' parameters and their uncertainties}
\footnotesize
\centering
\vspace{-3mm}
\begin{tabular}{lccccc}
\hline \hline
Model         & Search  & Best-fit & 1$\sigma$  & 3$\sigma$ \\ 
parameters    & range   &   values &(16--84\%)  & (0.14--99.86\%)\\ 
(1)           & (2)     & (3)      & (4)        & (5)\\ 
\hline\hline
\multicolumn{5}{c}{\skysampler}\\ 
\hline
\underline{Linear \ml$_{\rm F814W}$:} & \multicolumn{3}{c}{$\chi^2_{\rm red, min} \approx 0.75$}  & ~ \\
$\lg(M_{\rm BH}$/\Msun) &7$\rightarrow$11& 9.11 & $\pm$0.03 & $\pm$0.10 \\ 
$M/L_0$ (\Msun/\Lsun)   &0$\rightarrow$5& 3.31 & $\pm$0.05 & $\pm$0.14 \\  
$\alpha$ (\Msun/\Lsun\ per arcsec)& 0$\rightarrow$1& 0.11 & $\pm$0.02 & $\pm$0.05\\ [1mm]
\hline
\underline{Gaussian \ml$_{\rm F814W}$:} & \multicolumn{3}{c}{$\chi^2_{\rm red, min} \approx 0.75$}  & ~ \\
$\lg(M_{\rm BH}$/\Msun)  &7$\rightarrow$11& 9.10 & $\pm$0.07 & $\pm$0.20 \\ 
$M/L_0$ (\Msun/\Lsun)    &0$\rightarrow$5& 0.40 & $\pm$0.24 & $\pm$0.73 \\ 
$M/L_1$ (\Msun/\Lsun)    &0$\rightarrow$5& 3.13 & $\pm$0.16 & $\pm$0.49 \\ 
$\sigma_{\rm Gaussian}$ (arcsec)&0$\rightarrow$1& 0.93 & $\pm$0.33 & $\pm$0.98 \\[1mm] 
\hline
\underline{F555W MGE}: & \multicolumn{3}{c}{$\chi^2_{\rm red, min} \approx 0.72$}  & ~ \\
$\lg(M_{\rm BH}$/\Msun)  &7$\rightarrow$11& 9.15 & $\pm$0.05 & $\pm$0.15 \\ 
\ml$_{\rm F555W}$ (\Msun/\Lsun)    &0$\rightarrow$10& 7.02 & $\pm$0.11 & $\pm$0.34 \\ 
\hline\hline
\multicolumn{5}{c}{Gaussian}\\ 
\hline
\underline{Linear \ml$_{\rm F814W}$:}  & \multicolumn{3}{c}{$\chi^2_{\rm red, min} \approx 0.73$}  & ~ \\
$\lg(M_{\rm BH}$/\Msun)  &7$\rightarrow$11& 9.14 & $\pm$0.04 & $\pm$0.13 \\ 
$M/L_0$ (\Msun/\Lsun)    &0$\rightarrow$5& 2.88 & $\pm$0.11 & $\pm$0.33 \\ 
$\alpha$ (\Msun/\Lsun\ per arcsec)& 0$\rightarrow$1 & 0.49 & $\pm$0.10 & $\pm$0.30 \\ [1mm]
\hline
\underline{Gaussian \ml$_{\rm F814W}$:} & \multicolumn{3}{c}{$\chi^2_{\rm red, min} \approx 0.73$}  & ~ \\
$\lg(M_{\rm BH}$/\Msun)  &7$\rightarrow$11& 9.09 & $\pm$0.06 & $\pm$0.17 \\ 
$M/L_0$ (\Msun/\Lsun)    &0$\rightarrow$5& 0.04 & $\pm$0.09 & $\pm$0.20 \\ 
$M/L_1$ (\Msun/\Lsun)    &0$\rightarrow$5& 3.32 & $\pm$0.10 & $\pm$0.28 \\
$\sigma_{\rm Gaussian}$ (arcsec)&0$\rightarrow$1& 2.06 & $\pm$0.28 & $\pm$0.87 \\ 
\hline
\underline{F555W MGE}: & \multicolumn{3}{c}{$\chi^2_{\rm red, min} \approx 0.72$}  & ~ \\
$\lg(M_{\rm BH}$/\Msun)  &7$\rightarrow$11& 9.13 & $\pm$0.04 & $\pm$0.11 \\ 
\ml$_{\rm F555W}$ (\Msun/\Lsun)    &0$\rightarrow$10& 6.94 & $\pm$0.12 & $\pm$0.29 \\ 
\hline
\end{tabular}
\parbox[t]{0.47\textwidth}{\small \textbf{Notes:} In these \kinms\ models, we fixed all molecular gas and nuisance parameters at their best-fit values as of their default models listed in Section~\ref{tab:mcmc-results}.}
\label{tab:mcmc-results_mlvary}
\end{table}

\begin{figure}
    \includegraphics[width=\columnwidth]{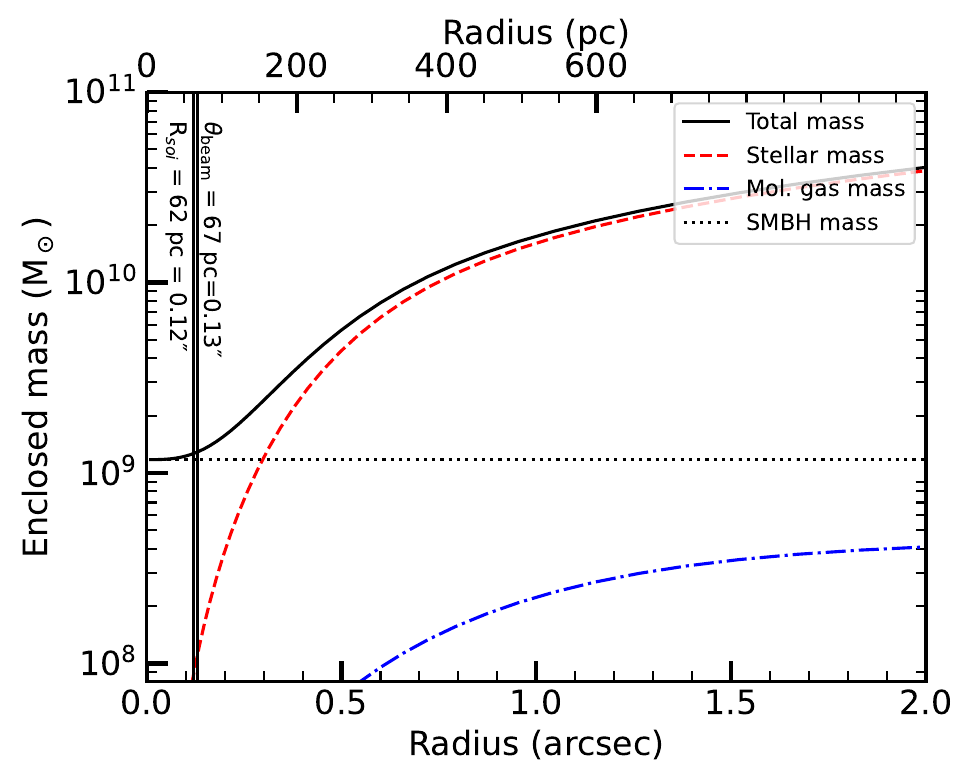}
    \caption{Enclosed mass of NGC 4061 (black solid line) as a function of radius, showing the contributions of all mass components: \Mbh, stars, and ISM.}
    \label{fig:NGC4061_cumulative_mass}
\end{figure}

\begin{figure*}
    \centering
    \includegraphics[width=\textwidth]{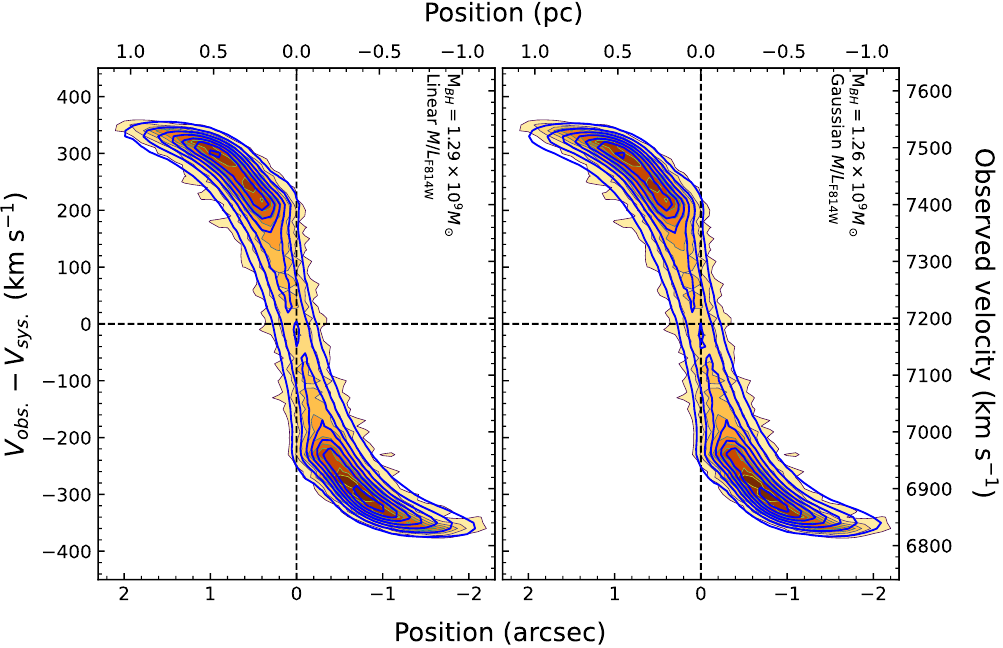}
    \caption{Position–velocity diagrams along the major axis for the best-fitting \kinms\ models, in which the \cotwo\ gas surface-brightness distribution is described using the \skysampler\ tool. Models adopting linear ({\it left}) and Gaussian ({\it right}) $\ml_{\rm F814W}(r)$ profiles are shown.}
    \label{fig:NGC4061_varyML_Skysampler}
\end{figure*}

\begin{figure}
    \centering
    \includegraphics[width=\columnwidth]{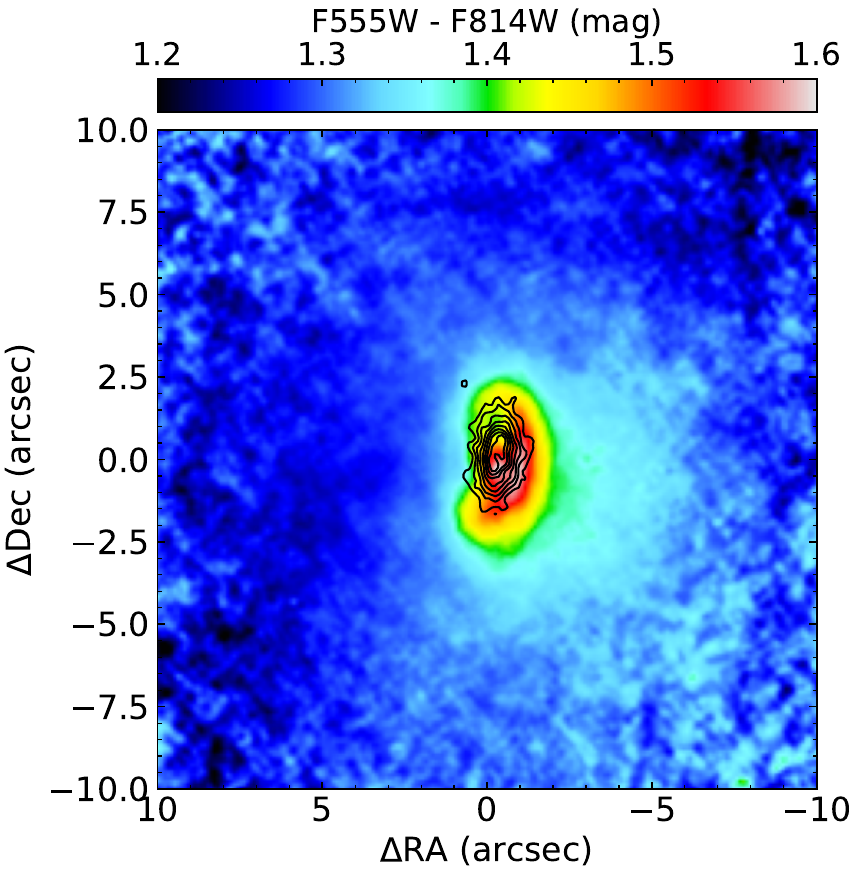}
    \caption{The \cotwo\ contours are overlaid on the nuclear F555W–F814W color map of NGC~4061 derived from \hst/WFPC2 images, showing their spatial coincidence with the prominent dust lane. The observed color variations are clearly dominated by dust extinction rather than by changes in the underlying stellar population.}
    \label{fig:F555W-F814W}
\end{figure}

\subsubsection{Thick Disk Assumption} \label{sec:uncer_thick_disc}

In our \kinms\ modeling, we assumed a geometrically thin \cotwo\ CND by fixing its vertical scale height to zero ($d=0\arcsec$). Physically, the molecular disk may have a finite vertical extent, so we tested the impact of this assumption by introducing a constant vertical thickness as an additional free parameter. We performed this test using two independent surface-brightness prescriptions: one based on the \skysampler\ tool and another using a Gaussian profile. In both cases, the best-fit parameters were consistent with those in Table~\ref{tab:mcmc-results}, differing by less than 5\%. The inferred vertical thicknesses were $d=0\farcs017\pm0\farcs003$ for the \skysampler\ model and $d=0\farcs025\pm0\farcs005$ for the Gaussian model. Since both values are smaller than the synthesized beam, the assumption of a razor-thin \cotwo\ disk is well justified for NGC~4061.

\subsubsection{Turbulent Velocity Dispersion of the Gas}\label{sec:uncer_sigma_vary}

Although our KinMS models assumed a constant turbulent velocity dispersion for the \cotwo\ CND, $\sigma_{\rm gas}$ may vary with both radius and azimuth across the disk. Moreover, beam smearing can artificially increase the central velocity dispersion, potentially overestimate the inferred \Mbh. To evaluate these effects on the \Mbh\ uncertainty, we parameterized $\sigma_{\rm gas}$ to several functional forms below. Here, we imposed a lower limit of $\sigma_{\rm gas,min} = 1$~\kms\ to prevent unrealistically narrow line profiles \citep{Barth16b}:

{\it Linearity:} $\sigma_{\rm gas}(r) = \alpha \times r + \beta$, where $a$ and $b$ are free parameters. The fits yielded $\alpha \approx 0$, with $\beta = 13.91$~\kms\ for the \skysampler\ model and $b = 15.51$~\kms\ for the Gaussian model. All other best-fitting KinMS parameters remain consistent with those from the constant-dispersion models presented in Section~\ref{sec:results} and Table~\ref{tab:mcmc-results}.

{\it Exponential:} $\sigma_{\rm gas}(r) = \sigma_0 \exp(-r/r_0) + \sigma_1$, where $\sigma_0$, $\sigma_1$, and $r_0$ are free parameters. The best-fitting \kinms\ model using the \skysampler\ tool yields \Mbh\ $= (1.23^{+0.22}_{-0.18}) \times 10^9$~\Msun\ and $M/L_{\rm F814W} = 3.44 \pm 0.08$~(\Msun/\Lsun), with $\sigma_0 = 83.45 \pm 3.27$~\kms, $\sigma_1 = 14.15 \pm 0.55$~\kms, and $r_0 = 0\farcs09 \pm 0\farcs04$. The corresponding model assuming a Gaussian gives \Mbh\ $= (1.20^{+0.21}_{-0.18}) \times 10^9$~\Msun\ and $M/L_{\rm F814W} = 3.37 \pm 0.07$~(\Msun/\Lsun), with $\sigma_0 = 65.98 \pm 1.14$~\kms, $\sigma_1 = 13.03 \pm 0.29$~\kms, and $r_0 = 0\farcs12 \pm 0\farcs04$.

{\it Gaussian:} $\sigma_{\rm gas}(r) = \sigma_0 \exp\left[-(r - r_0)^2 / 2\mu^2\right] + \sigma_1$, where $\sigma_0$, $\sigma_1$, $\mu$, and $r_0$ are free parameters. The parameter $r_0$ was allowed to vary over positive and negative values to account for potential offsets of the velocity-dispersion peak from the center.  The best-fitting \kinms\ model using the \skysampler\ tool yields \Mbh\ $= (1.20^{+0.18}_{-0.16}) \times 10^9$~\Msun\ and $M/L_{\rm F814W} = 3.47 \pm 0.06$~(\Msun/\Lsun), with $\sigma_0 = 69.63 \pm 2.87$~\kms, $\sigma_1 = 9.87 \pm 0.38$~\kms, $r_0 = 0\farcs01 \pm 0\farcs001$, and $\mu = 0\farcs01 \pm 0\farcs001$. The corresponding model assuming a Gaussian gas distribution gives \Mbh\ $= (1.23^{+0.15}_{-0.13}) \times 10^9$~\Msun\ and $M/L_{\rm F814W} = 3.37 \pm 0.07$~(\Msun/\Lsun), with $\sigma_0 = 71.04 \pm 1.22$~\kms, $\sigma_1 =13.49 \pm 1.35$~\kms, $r_0 = 0\farcs01 \pm 0\farcs001$, and $\mu = 0\farcs01 \pm 0\farcs001$.

These results indicate that assuming a constant $\sigma_{\rm gas}$ provides an adequate description of the \cotwo\ disk kinematics for \Mbh\ dynamical modeling. While the choice of radial parameterization for $\sigma_{\rm gas}(r)$ does influence the inferred \Mbh, the effect is modest: adopting a linear gradient yields negligible changes, and the resulting systematic uncertainties are $\lesssim$5\% and $\lesssim$5\% for the exponential and Gaussian $\sigma_{\rm gas}(r)$ profiles, respectively.

\subsubsection{Inclination} \label{sec:uncer_mge_deprojection}

In nearly face-on CNDs (inclination $\lesssim 30^\circ$), dynamical models often struggle to constrain both the SMBH mass and stellar $M/L$, leading to large uncertainties and asymmetric posterior distributions \citep[e.g.,][]{Smith19}. This limitation arises because the deprojection of the MGE into a 3D intrinsic stellar mass model becomes degenerate when the system is viewed close to face-on. In contrast, NGC~4061 exhibits a well-constrained \cotwo-CND inclination of $i \approx 60^\circ$, allowing for a unique 3D deprojection of the stellar mass distribution. Consequently, inclination-related uncertainties in our derived \Mbh\ and $M/L_{\rm F814W}$ are negligible (see Figure~\ref{fig:triangle} and Figure~\ref{fig:triangle1}).

\begin{figure*}
    \centering
    \includegraphics[width=\textwidth]{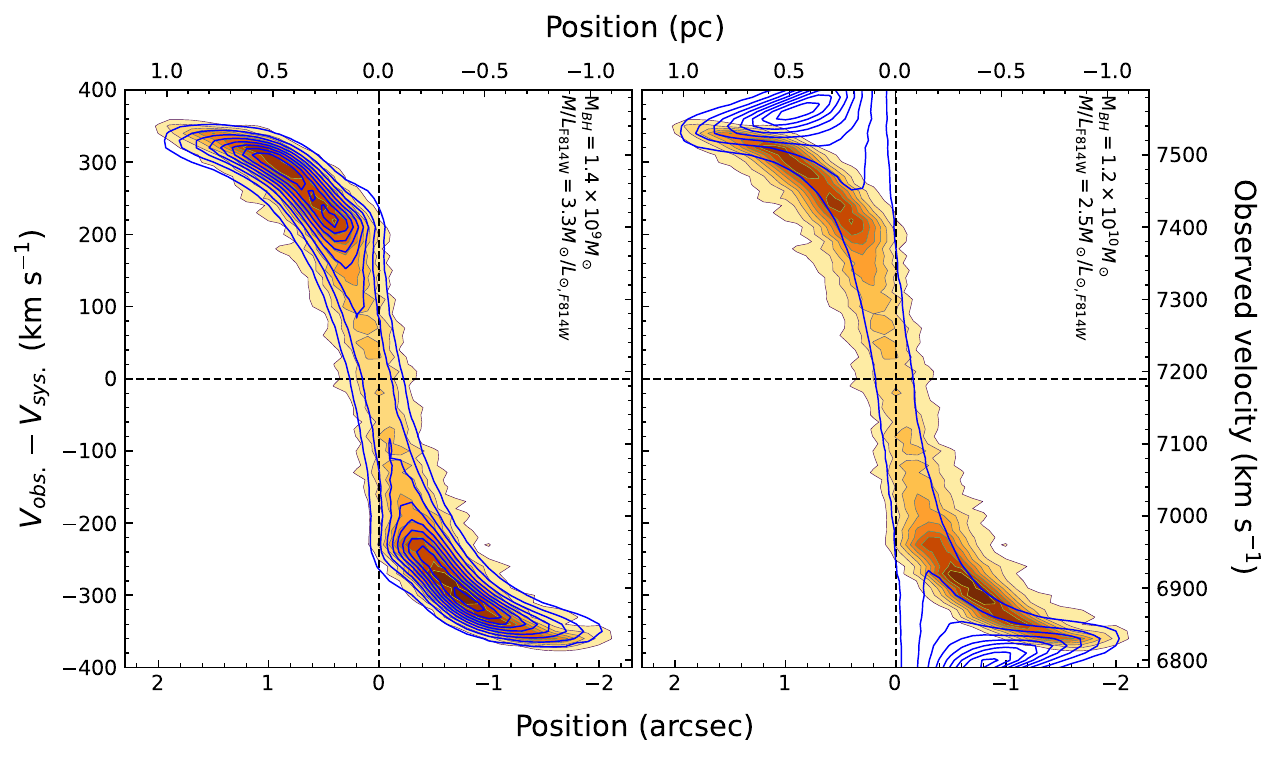}
    \caption{Position–velocity diagrams along the major axis for the \skysampler-based \kinms\ models are shown for two assumed SMBH masses calibrated following the \citet{Kormendy13} \Mbh--$\sigma$ correlations and a constant $\ml_{\rm F814W}$. The {\it left} panel adopts $M_{\rm BH}=2.0\times10^{9}$~\Msun, inferred from the central stellar velocity dispersion of $\sigma\approx290$~\kms\ measured by \citet{Pinkney2005}. The {\it right} panel adopts a substantially larger mass, $M_{\rm BH}=1.2\times10^{10}$~\Msun, corresponding to the high stellar velocity dispersion of $\sigma\approx459$~\kms\ listed in the Hypercat catalog. Both models fail to describe the \cotwo-CND's central rotation.}
    \label{fig:NGC4061_pvd_CompareBH}
\end{figure*}

\subsubsection{Mass-to-Light Ratio Model Variations} \label{sec:uncer_mass_vary}

Our fiducial analysis adopts a spatially constant \ml$_{\rm F814W}$. In reality, however, \ml$_{\rm F814W}$ may exhibit radial variations as a result of mass segregation \citep{Nguyen2025b}, potentially giving rise to a centrally enhanced profile that could partially mimic, or augment, the dynamical signature of a compact central dark mass. To investigate this effect, we constructed a set of test models using the same framework described in Section~\ref{sec:bayesian-infer}, but allowing for radially varying \ml$_{\rm F814W}$ profiles. Specifically, we considered two functional forms: (1) a linear profile, $M/L_{\rm F814W}(r) = M/L_0 + \alpha \times r$, and (2) a Gaussian profile, $M/L_{\rm F814W}(r) = M/L_0 \exp(-r^2 / 2\sigma_{\rm Gaussian}^2) + M/L_1$, where $M/L_0$ denotes the central mass-to-light ratio in both cases, $\alpha$ is the linear gradient, $M/L_1$ is a constant offset, and $\sigma_{\rm Gaussian}$ characterizes the Gaussian width.    For these experiments, all nuisance parameters and molecular gas properties were held fixed at their best-fitting values from the fiducial models (Table~\ref{tab:mcmc-results}). At the same time, the \Mbh\ was allowed to vary freely. The resulting \kinms\ models were analyzed within the same Bayesian framework, and the corresponding best-fit parameters are summarized in Table~\ref{tab:mcmc-results_mlvary}.

Figure~\ref{fig:NGC4061_varyML_Skysampler} compares the two preferred \kinms\ models that adopt the \skysampler\ prescription for the \cotwo\ surface brightness distribution, for each of the radially varying $M/L_{\rm F814W}(r)$ parameterizations. Both the linear and Gaussian $M/L_{\rm F814W}(r)$ profiles reproduce the observed \cotwo\ kinematics across the circumnuclear disk, despite the absence of independent observational evidence for color gradients or stellar population variations in the central region of NGC~4061. Relative to the fiducial models assuming a constant $M/L_{\rm F814W}$ (Section~\ref{sec:results}; Table~\ref{tab:mcmc-results}), the inferred \Mbh\ values differ by less than 10\% for the linear profile and less than 7\% for the Gaussian profile, remaining fully consistent within the quoted $1\sigma$ uncertainties. Consistent results are also obtained for \kinms\ models that assume a Gaussian functional form for the \cotwo\ surface brightness, for which the inferred \Mbh\ values differ by less than 17\% for the linear profile and less than 5\% for the Gaussian profile, and likewise agree within the $1\sigma$ uncertainties.

To further investigate the issue of possible spatial $M/L_{\rm F814W}$ variation, we constructed a F555W–F814W color map using an additional \hst/F555W image in combination with the F814W image (Figure~\ref{fig:F555W-F814W}). Before generating the color map, the astrometrically aligned images were cross-convolved with each other’s PSFs (i.e., the F555W image convolved with the F814W PSF and vice versa) to minimize artificial color gradients arising from differences in PSF width. We then subtracted the sky background from each image, measured within an annulus spanning (35–40)\arcsec from the galaxy center. The resulting color map reveals a spatial coincidence between the \cotwo\ contours and regions of enhanced F555W–F814W color, while the remainder of the nuclear region exhibits nearly uniform color. This correspondence indicates that the observed color variations are dominated by dust extinction rather than intrinsic changes in the stellar population.   

\begin{figure*}
    \centering
    \includegraphics[width=\columnwidth]{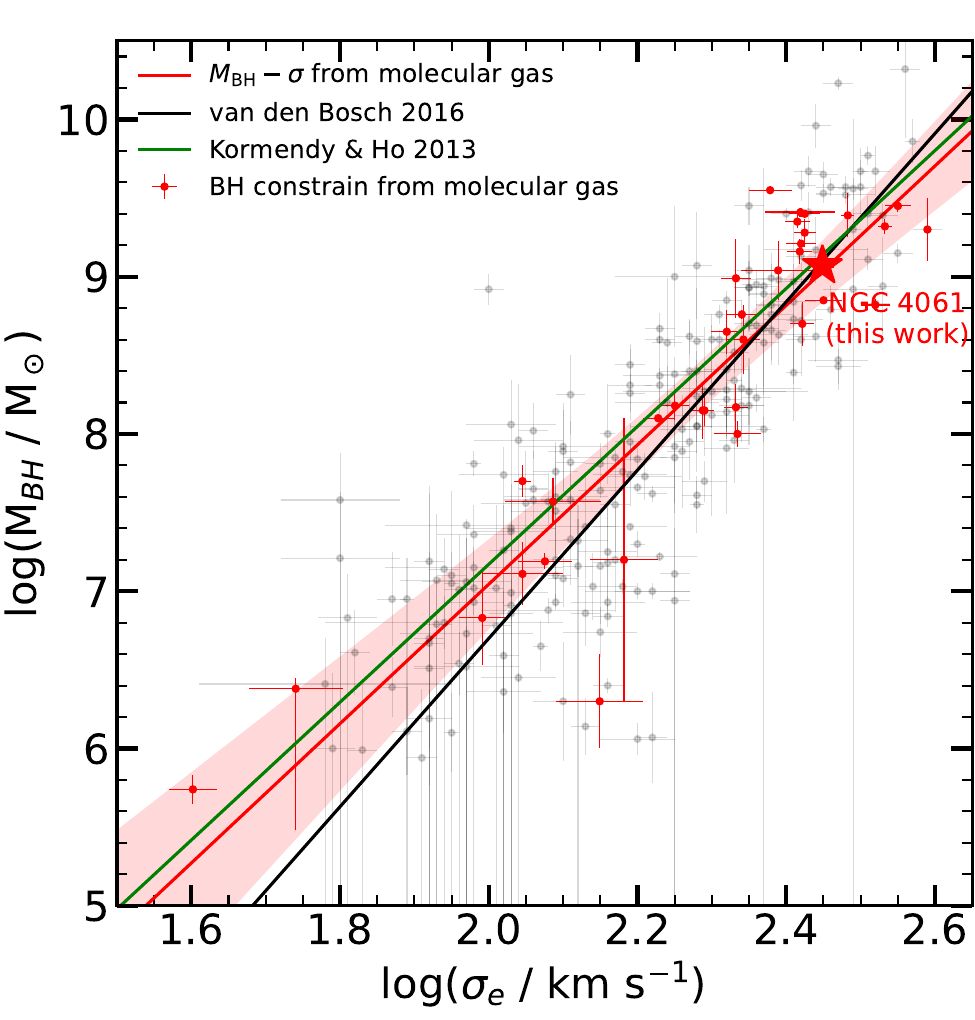} 
    \hspace{5mm}
    \includegraphics[width=\columnwidth]{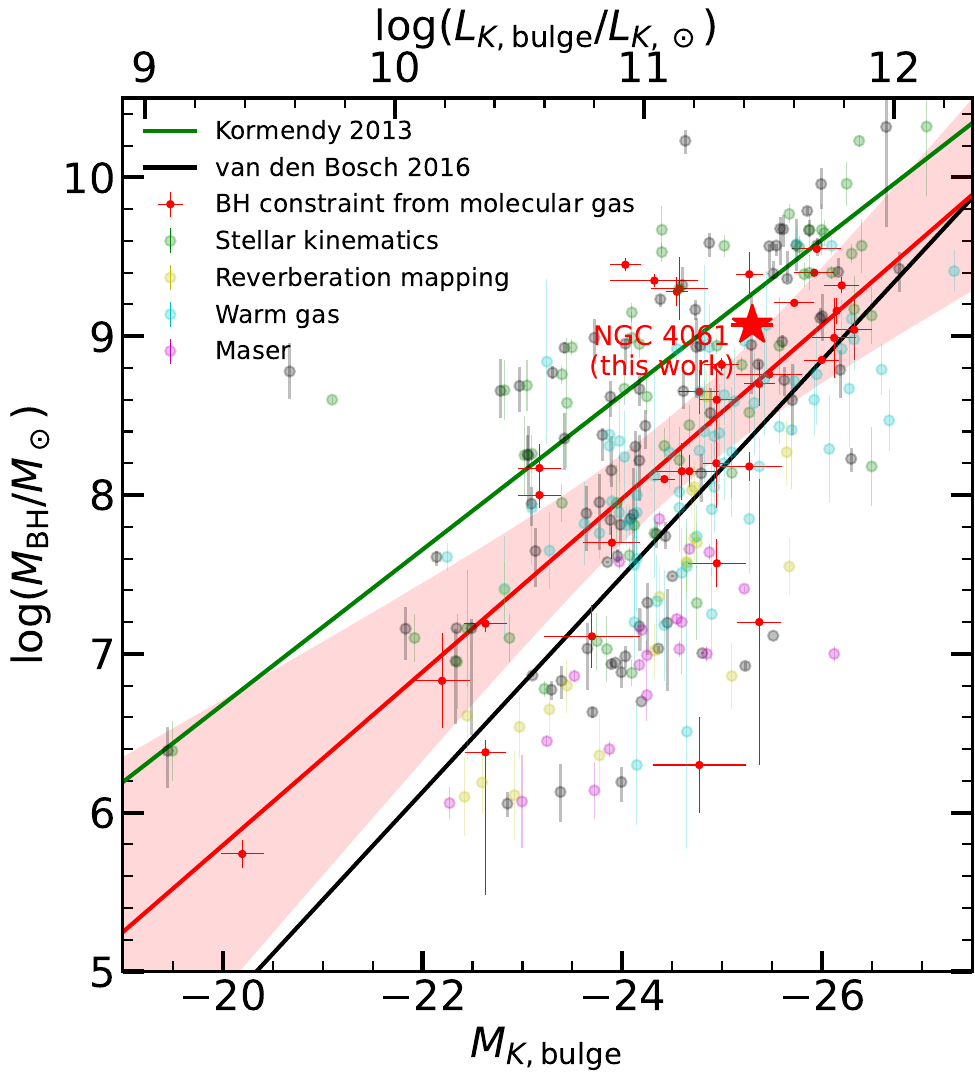}
    \caption{Our \Mbh\ measurement for NGC~4061 is shown in comparison with both the well-established literature \Mbh--$\sigma$ and \Mbh--$L_{K,\rm bulge}$ relations, as well as the molecular-gas-based relations alone, including their intrinsic scatters.}
    \label{fig:bhmass-sigma}
\end{figure*}

\begin{table*}[]
\centering
\caption{SMBH mass measurements derived solely from molecular gas dynamics. Here, $\sigma_e$ denotes the stellar velocity dispersion within the galaxy's effective radius, and $M_K$ is the absolute $K$-band magnitude obtained from the HyperLeda database. All quoted uncertainties correspond to the $1\sigma$ confidence level.}
\begin{tabular}{ccccccc}
\hline
No. & Galaxy & $\log(M_{\rm BH}/ M_\odot)$ & $\sigma_e$ (\kms)  & $M_{K, \rm bulge}$ (mag) &$\log(L_{K,\rm bulge}/ L_\odot)$& References \\
(1) & (2) & (3) & (4) & (5) & (6) & (7) \\
\hline
 1 & Fairall 49 & $8.20 \pm 0.28$ & --             & $-24.94 \pm 0.13$ & $11.29 \pm 0.06$ & \citet{Lelli2022} \\
 2 & NGC 315   & $9.32 \pm 0.05$ & $341 \pm 7$     & $-26.19 \pm 0.18$ & $11.79 \pm 0.08$ & \citet{Boizelle21} \\
\multirow{2}{*}{3}&\multirow{2}{*}{NGC 383}&\multirow{2}{*}{$9.55 \pm 0.02$}&\multirow{2}{*}{$239 \pm 16$}&\multirow{2}{*}{$-25.94 \pm 0.24$}&\multirow{2}{*}{$11.69 \pm 0.11$}&\citet{North19}\\ \cline{7-7}
   &           &                 &                 &                   &          & \citet{Zhang2025} \\
 4 & NGC 404   & $5.74 \pm 0.09$ & $40 \pm 3$      & $-20.20 \pm 0.22$ & $ 9.39 \pm 0.10$ & \citet{Davis20} \\
 5 & NGC 524   & $8.60 \pm 0.22$ & $220 \pm 11$    & $-24.94 \pm 0.18$ & $11.29 \pm 0.08$ & \citet{Smith19, Smith21} \\
 6 & NGC 613   & $7.57 \pm 0.15$ & $122 \pm 18$    & $-24.94 \pm 0.29$ & $11.29 \pm 0.13$ & \citet{Combes19} \\
 7 & NGC 997   & $8.99 \pm 0.25$ & $215 \pm 10$    & $-26.12 \pm 0.22$ & $11.76 \pm 0.10$ & \citet{Dominiak24} \\
 8 & NGC 1097  & $8.15 \pm 0.10$ & $195 \pm 4$     & $-24.67 \pm 0.26$ & $11.18 \pm 0.12$ & \citet{Onishi2015} \\
 9 & NGC 1275  & $9.04 \pm 0.19$ & $245 \pm 28$    & $-26.32 \pm 0.18$ & $11.84 \pm 0.08$ & \citet{Nagai19} \\
10 & NGC 1326  & $7.11 \pm 0.20$ & $111 \pm 14$    & $-23.69 \pm 0.48$ & $10.79 \pm 0.22$ & \citet{Combes19} \\
11 & NGC 1332  & $8.82 \pm 0.04$ & $331 \pm 15$    & $-25.00 \pm 0.18$ & $11.31 \pm 0.08$ & \citet{Barth16b} \\
12 & NGC 1365  & $6.30 \pm 0.30$ & $141 \pm 19$    & $-24.77 \pm 0.46$ & $11.12 \pm 0.12$ & \citet{Combes19} \\
13 & NGC 1380  & $8.17 \pm 0.15$ & $215 \pm 8$     & $-23.17 \pm 0.22$ & $10.58 \pm 0.10$ & \citet{Kabasares22} \\
14 & NGC 1566  & $6.83 \pm 0.30$ & $98 \pm 7$      & $-22.19 \pm 0.28$ & $10.19 \pm 0.13$ & \citet{Combes19} \\
15 & NGC 1574  & $8.00 \pm 0.08$ & $216 \pm 16$    & $-23.17 \pm 0.22$ & $10.58 \pm 0.10$ & \citet{Ruffa23} \\
16 & NGC 1672  & $7.70 \pm 0.10$ & $111 \pm 3$     & $-23.89 \pm 0.29$ & $10.87 \pm 0.13$ & \citet{Combes19} \\
17 & NGC 1684  & $9.16 \pm 0.08$ & $262 \pm 10$    & $-26.14 \pm 0.00$ & $11.77 \pm 0.10$ & \citet{Dominiak24} \\
18 & NGC 3258  & $9.35 \pm 0.04$ & $260 \pm 10$    & $-24.32 \pm 0.44$ & $11.04 \pm 0.20$ & \citet{Boizelle19} \\
19 & NGC 3504  & $7.19 \pm 0.05$ & $119 \pm 10$    & $-22.63 \pm 0.22$ & $10.36 \pm 0.10$ & \citet{Nguyen20} \\
20 & NGC 3557  & $8.85 \pm 0.02$ & $282 \pm 16$    & $-26.00 \pm 0.18$ & $11.71 \pm 0.08$ & \citet{Ruffa19} \\
21 & NGC 3593  & $6.38^{+0.08}_{-0.90}$ & $55 \pm 7$& $-22.63 \pm 0.22$& $10.36 \pm 0.10$ & \citet{Nguyen22} \\
22 & NGC 3665  & $8.76 \pm 0.03$ & $219 \pm 10$    & $-25.47 \pm 0.33$ & $11.50 \pm 0.15$ & \citet{Onishi17} \\
23 & NGC 4061  & $9.07 \pm 0.04$ & $290 \pm ...$     & $-25.27 \pm 0.13$ & $11.60 \pm 0.09$ & This Work \\
24 & NGC 4261  & $9.21 \pm 0.01$ & $263 \pm 12$    & $-25.72 \pm 0.20$ & $11.60 \pm 0.09$ & \citet{Ruffa23} \\
25 & NGC 4429  & $8.18 \pm 0.09$ & $178 \pm 8$     & $-25.27 \pm 0.33$ & $11.42 \pm 0.15$ & \citet{Davis18} \\
26 & NGC 4526  & $8.65 \pm 0.14$ & $209 \pm 10$    & $-24.77 \pm 0.20$ & $11.12 \pm 0.09$ & \citet{Davis13} \\
27 & NGC 4697  & $8.10 \pm 0.02$ & $169 \pm 8$     & $-24.42 \pm 0.11$ & $11.08 \pm 0.05$ & \citet{Davis17} \\
28 & NGC 4751  & $9.45 \pm 0.04$ & $355 \pm 14$    & $-24.03 \pm 0.15$ & $10.92 \pm 0.07$ & \citet{Dominiak2024b} \\
29 & NGC 4786  & $8.70 \pm 0.14$ & $264 \pm 10$    & $-25.37 \pm 0.15$ & $11.46 \pm 0.07$ & \citet{Kabasares2024} \\
30 & NGC 5193  & $8.15 \pm 0.18$ & $194 \pm 7$     & $-24.59 \pm 0.26$ & $11.15 \pm 0.12$ & \citet{Kabasares2024} \\
31 & NGC 6861  & $9.30 \pm 0.20$ & $389 \pm 3$     & $-24.57 \pm 0.29$ & $11.14 \pm 0.13$ & \citet{Kabasares22} \\
\multirow{2}{*}{32}&\multirow{2}{*}{NGC 7052}&\multirow{2}{*}{$9.40 \pm 0.02$}&\multirow{2}{*}{$266 \pm 13$}&\multirow{2}{*}{$-25.92 \pm 0.20$}&\multirow{2}{*}{$11.68 \pm 0.09$}&\citet{Smith19, Smith21}\\ \cline{7-7}
   &           &                 &                 &                   &           & \citet{Ngo2025a} \\
33 & NGC 7469  & $7.20 \pm 0.90$ & $152 \pm 16$    & $-25.37 \pm 0.22$ & $11.46 \pm 0.10$ & \citet{Nguyen21} \\
34 & PGC 11179 & $9.28 \pm 0.09$ & $266 \pm 9$     & $-24.55 \pm 0.11$ & $11.13 \pm 0.05$ & \citet{Cohn2023} \\
35 & UGC 2698  & $9.39 \pm 0.14$ & $304 \pm 6$     & $-25.27 \pm 0.13$ & $11.42 \pm 0.06$ & \citet{Cohn21} \\
\hline
\end{tabular}
\label{tab:mbh-compile}
\end{table*}

These findings demonstrate that our ALMA-based measurement of \Mbh\ is only weakly sensitive to the assumed stellar mass-to-light ratio profile, including plausible variations driven by dust attenuation or stellar population changes across the \cotwo\ CND. This insensitivity reflects the dominance of the central black hole over the gravitational potential within the inner $0\farcs2$. We therefore conclude that uncertainties associated with the stellar mass model contribute at the $\sim$10\% level to the total error budget of \Mbh.

\subsubsection{F814W MGE Stellar Mass Model without Dust Masking}\label{sec:f815w_mge_nomask}

In constructing the fiducial \hst/WFPC2 F814W MGE stellar mass model, we masked the prominent central dust lane visible in the \hst\ image. However, the total molecular gas mass inferred from the \cotwo\ circumnuclear disk (extending to $r \lesssim 2\arcsec$) is nearly two orders of magnitude smaller than the enclosed stellar mass within the same radius (Figure~\ref{fig:NGC4061_cumulative_mass}). To assess the potential impact of dust obscuration on our SMBH mass measurement, we repeated the MGE modeling without applying a dust mask.   The resulting dust-unmasked \hst/WFPC2 F814W MGE model was then adopted as the stellar mass input for both sets of \kinms\ models. Using the \skysampler\ prescription for the CND surface brightness, we obtained \Mbh\ $=(7.41 \pm 1.92)\times10^8$~\Msun\ and $M/L_{\rm F814W}=5.50\pm0.16$~(\Msun/\Lsun). The corresponding model employing a simple Gaussian surface brightness profile yields \Mbh\ $=(6.92\pm1.16)\times10^8$~\Msun\ and $M/L_{\rm F814W}=5.42\pm0.06$~(\Msun/\Lsun).  The close agreement between these results and those from the dust-masked models demonstrates that dust attenuation has a negligible impact on our dynamical determination of \Mbh.

\subsubsection{Using the Alternative F555W MGE Stellar Mass Model} \label{sec:f555w_mge}

We assessed the effect of adopting an alternative stellar mass model derived from the \hst/WFPC2 F555W image on our dynamical determination of \Mbh. In this test, we replaced the fiducial \hst/WFPC2 F814W MGE model described in Section~\ref{sec:stellar-mass} with the corresponding F555W-based MGE model. For consistency, we adopted a photometric zero point of 24.683~mag and a solar absolute magnitude of 4.82~mag \citep{Willmer18} in the F555W band. This alternative stellar mass model was then implemented within the \kinms\ framework, considering both the \skysampler\ and Gaussian prescriptions for the \cotwo\ CND's surface brightness. In these tests, only \Mbh\ and the stellar \ml$_{\rm F555W}$ were treated as free parameters, while all remaining model parameters were fixed to their best-fitting values from the fiducial models (Table~\ref{tab:mcmc-results}). The resulting best-fit values of \Mbh\ and \ml$_{\rm F555W}$ for both modeling approaches are summarized in Table~\ref{tab:mcmc-results_mlvary}.  Overall, adopting the F555W-based stellar mass model introduces an additional uncertainty of approximately 20\% in the inferred SMBH mass for NGC~4061, likely reflecting the greater sensitivity of the F555W band to dust attenuation compared to the F814W band.

\subsection{The Reliability of Our Measurements} \label{sec:previous-work}

Following the argument of \citet{Rusli2013a}, the impact of observational angular resolution on the precision of \Mbh\ measurements can be quantified by the resolving power $\zeta = 2r_{\rm soi}/\theta_{\rm FWHM}$, where $r_{\rm soi}$ is the SMBH SOI radius and $\theta_{\rm FWHM}$ is the synthesized beam size. Observations with $\zeta \geq 2$ (i.e., $\theta_{\rm FWHM} \leq r_{\rm soi}$) are generally sufficient to yield precise \Mbh\ determinations \citep[e.g.,][]{North19, Boizelle21}. For $\zeta < 2$, the inferred \Mbh\ values remain useful \citep[e.g.,][]{Davis14, Nguyen20}, but typically exhibit larger uncertainties and increased sensitivity to systematics, including uncertainties in the stellar mass profile and beam-smearing effects \citep{Rusli2013b, Barth16b, Nguyen21}. In the case of NGC~4061, our ALMA data correspond to $\zeta \approx 2$, indicating that the dynamical-\Mbh\ measurement presented in this work is robust.

As discussed in Section~\ref{sec:intro}, previous estimates of the SMBH mass in NGC~4061 relied solely on the \Mbh–$\sigma$ relation \citep[e.g.,][]{Kormendy13}, using low–spatial-resolution measurements of the velocity dispersion of either the stellar \citep[$\sigma \approx 290$~\kms;][]{Pinkney2005} or ionized gas components ($\sigma \approx 459$~\kms; Hypercat catalog). The ionized-gas velocity dispersion, in particular, appears to be unreliable. We compared our combined ALMA observations with \skysampler-based \kinms\ models adopting SMBH masses of $M_{\rm BH} \approx 2.0\times10^{9}$~\Msun\ (from the stellar dispersion) and $M_{\rm BH} \approx 1.2\times10^{10}$~\Msun\ (from the ionized-gas dispersion). In these models, all other parameters were fixed to the best-fit values listed in Table~\ref{tab:mcmc-results}, while only \ml$_{\rm F814W}$ was allowed to vary to constrain the rotation of the outer \cotwo\ CND.  Figure~\ref{fig:NGC4061_pvd_CompareBH} shows that neither model provides an adequate fit to the ALMA data. Although their favored \ml$_{\rm F814W}$ values decrease with increasing BH mass, both models systematically overpredict the rotation of the \cotwo\ CND—either within the inner $0\farcs6$ for $M_{\rm BH} = 2.0\times10^{9}$~\Msun\ or across the entire disk for $M_{\rm BH} \approx 1.2\times10^{10}$~\Msun. These results support our molecular-gas dynamical measurement and indicate that the SMBH mass in NGC~4061 is of order $10^{9}$~\Msun, at most slightly above one billion solar masses.

We also should note that a $\sim$$10^{9}$~\Msun\ SMBH at $\sim$107 Mpc has an event-horizon angular size of only $\sim$0.4~$\mu$as, far below the $\sim$10~$\mu$as reach of ngEHT \citep[e.g.,][]{Johnson2023, Shlentsova2024}, making direct horizon-scale imaging impossible at that distance.

\subsection{NGC~4061 Positions on the Molecular Gas Alone \Mbh--$\sigma$ and  \Mbh--$M_\star$ Scaling Relations} \label{sec:Mbh-sigma}

Our SMBH mass estimate for NGC~4061 is consistent within the $+1\sigma$ scatter of the \Mbh--$\sigma$ relations compiled by \citet{Bosch2016} and \citet{Kormendy13}, as shown in Figure~\ref{fig:bhmass-sigma}. This agreement is expected, as both relations are calibrated primarily using classical bulges and elliptical galaxies, including brightest cluster galaxies, a category to which NGC~4061 belongs. The \Mbh\ values predicted by these relations are $2.0 \times 10^9$~\Msun.

In contrast, on the \Mbh--$L_{K,\rm bulge}$ plane, our derived \Mbh\ for NGC~4061 is consistent within $-1\sigma$ of the compilation by \citet{Kormendy13}, but lies nearly $+2\sigma$ above the relation presented by \citet{Bosch2016}. This discrepancy likely reflects differences in the underlying datasets and measurement techniques adopted in the two compilations. Specifically, \citet{Kormendy13} included SMBH masses derived from stellar kinematics, ionized gas dynamics, and maser measurements, whereas \citet{Bosch2016} additionally incorporated reverberation-mapping estimates.

Over the past 12 years since the first successful molecular-gas-based \Mbh\ measurement, obtained using Combined Array for Research in Millimeter-Wave Astronomy (CARMA) observations of NGC~4526 by \citet{Davis13}, an additional 34 SMBH masses have been measured using ALMA. These molecular-gas-based measurements now span a mass range from $5\times10^5$ to $3.5\times10^9$~\Msun, including the measurement for NGC~4061 presented here. We summarize all 35 molecular-gas-based \Mbh\ measurements in Table~\ref{tab:mbh-compile}.

We then recompiled these scaling relations using only the 35 molecular-gas-based \Mbh\ measurements by performing linear regression analyses with the \textsc{linregress} routine from the \textsc{SciPy} package \citep{Virtanen2020}. Galaxy $K$-band magnitudes were obtained from the HyperLeda database\footnote{\url{http://atlas.obs-hp.fr/hyperleda/}}. All regressions were performed in log–linear space, with all quantities converted to logarithmic units before fitting, and a flat prior for fitting parameters was also applied. The resulting best-fit relations, including the 95\% confidence-level uncertainties, shown as red lines in Figure~\ref{fig:bhmass-sigma}, yield

\begin{equation*}
    \log\left(\frac{M_{\rm BH}}{M_\odot}\right) = -(1.25 \pm 0.94) + (4.20 \pm 0.38)\log \left( \frac{\sigma_e}{\rm km\ s^{-1}}\right), 
\end{equation*}
and
\begin{equation*}
    \log\left(\frac{M_{\rm BH}}{M_\odot}\right) = -(4.46 \pm 0.30) - (0.52 \pm 0.09) \times M_{K,{\rm bulge}}
\end{equation*}

The molecular-gas-constrained \Mbh--$\sigma$ relation appears broadly consistent with the relation presented by \citet{Kormendy13}, differing only by a small, approximately constant negative offset in \Mbh\ ($\approx$0.1 dex). Because molecular-gas-based measurements rely on dynamically cold, well-settled CNDs, which are relatively insensitive to inflows, outflows, and dark matter contributions on these scales, the resulting \Mbh\ estimates are arguably more robust. The presence of only a small systematic offset likely reflects differences in modeling assumptions, thereby underscoring the overall consistency and reliability of the dynamical measurement techniques. In contrast, the discrepancy with the \citet{Bosch2016} relation can be attributed to the inclusion of the intrinsically less precise reverberation-mapping mass estimates in that compilation.

The molecular-gas-based \Mbh--$L_{K,\rm bulge}$ relation derived in this work exhibits a slope that is closely aligned with that of the relation compiled by \citet{Kormendy13}, resulting in two nearly parallel relations with a modest offset in normalization. This similarity suggests that the underlying scaling between \Mbh\ and bulge stellar mass is robust when restricted to samples dominated by classical bulges and early-type galaxies and calibrated using direct dynamical measurements. In contrast, the relation presented by \citet{Bosch2016} exhibits a noticeably different slope, likely reflecting the broader, more heterogeneous nature of their sample, which includes a substantial fraction of SMBH masses derived from reverberation mapping. The inclusion of indirect mass estimators, combined with differences in galaxy morphology and regression methodology, can modify both the slope and normalization of the resulting scaling relation, thereby accounting for the observed divergence relative to the purely dynamical and molecular-gas-based relations.

\section{Conclusions}\label{sec:conclusion}

We derive the first dynamical SMBH mass in NGC~4061 through dynamical modeling of cold molecular gas, using ALMA \cotwo\ observations that combine medium–angular-resolution data from Cycle~7 ($\theta_{\rm beam} \approx 0\farcs32 \times 0\farcs24$) with higher–resolution archival data from Cycle~6 ($\theta_{\rm beam} \approx 0\farcs11 \times 0\farcs12$). The resulting data set attains a synthesized beam of $0\farcs16 \times 0\farcs13$, corresponding to $83.2 \times 67.6$~pc$^{2}$, which closely matches the estimated SOI radius of the SMBH ($r_{\rm soi} \approx 0\farcs12$).   By exploring multiple prescriptions for the spatial distribution of the molecular gas in the \cotwo\ CND, we obtain consistent constraints on the black hole mass, the stellar mass-to-light ratio, and other disk parameters. These models yield a robust dynamical measurement of $M_{\rm BH} = (1.17^{+0.08}_{-0.10}\,[{\rm stat.}] \pm 0.43\,[{\rm syst.}]) \times 10^{9}$~\Msun\ and an $I$-band stellar mass-to-light ratio of $M/L_{\rm F814W} = 3.46^{+0.07}_{-0.06}\,[{\rm stat.}] \pm 0.10\,[{\rm syst.}]$~\Msun/\Lsun. Our results highlight the importance of high–spatial-resolution ALMA observations of cold molecular gas for precision SMBH mass measurements. Provided that the synthesized beam remains comparable to the BH’s SOI, \cotwo\ serves as a reliable dynamical tracer, in contrast to warm ionized gas, which is frequently affected by non-circular motions and other dynamical disturbances.

Our results conclusively resolve the long-standing discrepancy among previous indirect \Mbh-estimates for NGC~4061, arising from inconsistent stellar velocity dispersion measurements, and demonstrate that the exceptionally large dispersion reported in the literature is likely incorrect.

The \Mbh--$\sigma$ and \Mbh--$L_{K,\rm bulge}$ relations derived from molecular-gas-based measurements alone are consistent with the relations of \citet{Kormendy13}, but not with those of \citet{Bosch2016}. This likely reflects the use of direct dynamical SMBH mass measurements in the former, compared to the inclusion of reverberation-mapping estimates and a more heterogeneous sample in the latter.

\section*{Acknowledgements}
This paper makes use of the following ALMA data: ADS/JAO.ALMA\#2018.1.00397.S and \#2019.1.00036.S. ALMA is a partnership of ESO (representing its member states), NSF (USA), and NINS (Japan), together with NRC (Canada) and NSC and ASIAA (Taiwan), and KASI (Republic of Korea), in cooperation with the Republic of Chile. The Joint ALMA Observatory is operated by ESO, AUI/NRAO, and NAOJ. The National Radio Astronomy Observatory is a facility of the National Science Foundation operated under cooperative agreement by Associated Universities, Inc.

\facility{ALMA and \hst/WFPC2}

\software{{\tt Python~3.12} \citep{VanRossum2009}, 
{\tt Matplotlib~3.6} \citep{Hunter2007}, 
{\tt NumPy~1.22} \citep{Harris2020}, 
{\tt SciPy~1.3} \citep{Virtanen2020},  
{\tt photutils~0.7} \citep{bradley2024}, 
{\tt AstroPy~5.1} \citep{AstropyCollaboration2022}, 
{\tt AdaMet 2.0} \citep{Cappellari2013a}, 
{\tt JamPy~7.2} \citep{Cappellari2020}, and
{\tt MgeFit~5.0} \citep{Cappellari2002}.
}

\bibliographystyle{aasjournalv7}
\bibliography{ngc4061}

\label{lastpage}
\end{document}